\documentclass[a4paper,fleqn,usenatbib]{mnras}

\usepackage[T1]{fontenc}
\usepackage{ae,aecompl}

\usepackage{graphicx}	
\usepackage{amsmath}	
\usepackage{amssymb}	
\usepackage{multicol}        
\usepackage{bm}		
\usepackage{color}
\usepackage{xcolor}

\title[{\it TESS} observations of Be stars]
{{\it TESS} observations of Be stars: a new interpretation}
\author[L. A. Balona and D. Ozuyar]{\
L. A. Balona$^{1}$\thanks{E-mail: lab@saao.ac.za} and
D. Ozuyar$^{2}$
\\
$^1$ South African Astronomical Observatory, 
P.O. Box 9, Observatory 7935, South Africa\\
$^2$ Ankara University, Faculty of Science, 
Dept. of Astronomy and Space Sciences, 06100, Tandogan, Ankara, Turkey}

\begin{document}

\date{Accepted .... Received ...}

\pagerange{\pageref{firstpage}--\pageref{lastpage}} \pubyear{2011}

\maketitle

\label{firstpage}

\begin{abstract}
Light curves of 57 classical Be stars in {\it TESS} sectors 1--15 are
examined.  In most Be stars, the periodogram shows groups at a fundamental and 
one or more harmonics, which we attribute to rotation. In about 40\,percent of 
the stars, the group is just a single narrow or slightly broadened peak.  In 
about 30\,percent, it consists of a multiple, closely spaced peaks.  These 
groups can be interpreted as non-coherent variations most likely associated 
with photospheric gas clouds.  Approximate rotational frequencies for about 
75\,percent of the stars can be derived.  Comparison with the projected 
rotational velocities shows that the photometric frequency is consistent with 
rotation. The first harmonic plays a prominent role in many Be stars and 
manifests itself in either single-wave or double-wave light curves.  The 
reduction in amplitude of $\beta$~Cep pulsations in a few Be stars during an 
outburst and their subsequent recovery is most likely an obscuration effect.  
Other instances of possible obscuration of the photosphere are suspected.  A 
simple model, which attempts to explain these observations and other general
properties of Be stars, is proposed.
\end{abstract}


\section{Introduction}

The classical Be stars are dwarf and giant B stars that show, or have
shown at some time, emission in the core of some Balmer lines (particularly 
the H$\alpha$ line) \citep{Slettebak1979, Porter2003}.  Stars in which the 
emission is a result of binary interaction are excluded from the definition.  
The emission is thought to be a result of mass loss from the star which is 
greatly assisted by rapid rotation.  

In many Be stars, spectroscopic line profile and light variations 
characteristic of non-radial pulsation (NRP) are observed with periods of 
about one day. For this reason it is generally believed that NRP acts as the 
trigger for mass-loss outbursts which occur from time to time 
\citep{Rivinius2013}.  Because the additional velocity provided by NRP is 
relatively small, this mechanism requires the star to be rotating in excess 
of about 90\,percent of the critical rotational velocity at the equator.  
In this hypothesis, every Be star must be rotating very near critical
velocity.

However, it has been shown that, as a group, Be stars rotate at rates which 
are well below the critical rotation rate.  \citet{Cranmer2005} found that 
early-type Be stars have an approximately uniform spread of intrinsic rotation 
speed that extends from 0.4--0.6 of critical, though a few may be rotating near
critical velocity.  Late-type Be stars exhibit progressively narrower ranges of
rotation speed as the effective temperature decreases. The lower limit rises to
reach critical rotation for the coolest Be stars.  More recently, 
\citet{Zorec2016} arrived at the same conclusion.  They found that in most Be 
stars the ratio of equatorial rotational velocity relative to the critical 
rotational velocity is $v/v_c \approx 0.65$. This ratio is characterized by a 
wide range $0.3 < v/v_c < 0.95$, suggesting that the probability that all Be 
stars are critical rotators is extremely low (see also \citealt{Cochetti2019}).  

In this paper we take the view that NRP, if present, is incidental and cannot 
play the main role in the mass loss mechanism.  An alternative mechanism needs 
to be found.

It is generally accepted that early-type stars have radiative envelopes
which cannot sustain magnetic fields.  Therefore photospheric activity cannot 
exist.  This has always been an important motivation for NRP as the source of 
mass loss in Be stars.  However, recent space photometry shows that a 
significant fraction of A and B stars have periods which are indistinguishable 
from the rotational periods \citep{Balona2013c,Balona2016a, Balona2017a}.  A 
recent study  concludes that about 35\,percent of all mid- to late-B stars 
exhibit rotational modulation \citep{Balona2019c}.  

It therefore appears that starspots may be present in many early-type stars.  
We may presume starspots are sources of activity which may lead to mass loss in 
rapidly rotating stars.  In this scenario, quasi-periodic line profile and 
light variation may possibly arise not only from starspots, but perhaps also 
by the material ejected from these active regions.

The quasi-periodicity of many Be stars tends to cluster in two frequency
groups, one twice the frequency of the other.  An example of this tendency can 
be found in ground-based photometry of several Be stars in the open cluster 
NGC\,3766 \citep{Balona1991b}. In some seasons the fundamental group
dominates and the light curve is a single wave.  At other seasons the first
harmonic group dominates and the light curve has a double-wave form. 

Photometric observations from space have confirmed this general pattern.  
Examples from the {\it MOST} satellite include HD\,163868 \citep{Walker2005b}, 
HD\,127756 and HD\,217543 \citep{Cameron2008}.  There are several examples 
from {\it CoRoT} as well \citep{Neiner2009,Semaan2018}.  The same is found in 
48\,Lib observed by the {\it STEREO} satellite \citep{Ozuyar2018} as do two of 
the three Be stars observed by {\it Kepler} \citep{Rivinius2016}.  These 
frequency groups have been interpreted as due to a large number of g-modes
driven by the $\kappa$ opacity mechanism.  They can perhaps also be regarded as
incoherent variations leading to quasi-periodicity.

\begin{table*}
\caption{List of Be stars selected for analysis.  The {\it TESS} catalogue
number, TIC, and the HD number are listed.  This is followed by the
variability classification based purely on the {\it TESS} light curve.
The number of {\it TESS} sectors is given by $N$.  The $V$ magnitude, the 
effective temperature, $T_{\rm eff}$ (K), and the literature reference for 
$T_{\rm eff}$ are given.  The stellar luminosity determined from the {\it GAIA} 
parallax is shown and is followed by the projected rotational velocity, 
$v\sin i$ (km\,s$^{-1}$).  The last column gives the spectral classification.} 
\label{data}
\resizebox{0.99\textwidth}{!}{
\begin{tabular}{rllrrrrrrrrl}
\hline
\multicolumn{1}{c}{TIC}                         & 
\multicolumn{1}{c}{HD}                          & 
\multicolumn{1}{c}{Var Type}                    & 
\multicolumn{1}{c}{N}                           & 
\multicolumn{1}{c}{$V$}                         & 
\multicolumn{1}{c}{$T_{\rm eff}$}               & 
\multicolumn{1}{c}{Ref}                         & 
\multicolumn{1}{c}{$\log \tfrac{L}{L_\odot}$}   & 
\multicolumn{1}{c}{$\nu_{\rm rot}$}             & 
\multicolumn{1}{c}{$v$}                         &
\multicolumn{1}{c}{$v\sin\,i$}                  &
\multicolumn{1}{c}{Sp Type}                     \\
\hline
  23037766 & 57150     &  ROT             &  1  &  4.670 & 22000 & 2 & 3.73 & 1.07  & 270 & 210 & B2V + B3IVne        \\
  42360166 & 191610    &  ROT             &  2  &  4.930 & 20470 & 9 & 3.34 & 1.5   & 280 & 320 & B2.5III             \\                              
  47296054 & 214748    &  ROT             &  1  &  4.180 & 11500 & 2 & 2.69 & 0.836 & 240 & 180 & B8/9IV/V            \\
  52665242 & 47054     &  ROT             &  1  &  5.520 & 13520 & 1 & 2.91 & 1.145 & 300 & 260 & B8IV/Ve shell       \\
  53992511 & 209522    &  SPB+ROT         &  1  &  5.952 & 22570 & 1 & 3.51 & 1.6   & 300 & 280 & B3V + B:            \\
  55295028 & 33599     &  ROT+BCEP        & 13  &  8.970 & 23200 & 7 & 4.46 & 0.905 & 480 & 200 & B3p shell           \\
  56179720 & 30076     &  SPB             &  1  &  5.810 & 26190 & 1 & 4.67 & -     & -   & 180 & B1V?e               \\
  65803653 & 56014     &  BCEP+ROT        &  1  &  4.650 & 24380 & 1 & 4.16 & 1.3   & 440 & 210 & B2V(e) shell        \\
  67251066 & 198183    &  ROT             &  1  &  4.540 & 15330 & 1 & 3.25 & 1.1   & 330 & 120 & B7.5IV              \\                              
  71132174 & 28497     &  SPB+ROT         &  1  &  5.410 & 32310 & 9 & 4.42 & 1.4   & 370 & 230 & B1Ve                \\
  71727949 & 41335     &  -               &  1  &  5.210 & 22500 & 2 & 4.25 & -     & -   & 340 & B1:IV/V:nne shell   \\
  75047606 & 79621     &  ROT             &  2  &  5.920 & 11710 & 1 & 2.31 & 1.67  & 290 & 170 & B9V                 \\                              
  81584371 & 54309     &  ROT             &  1  &  5.830 & 26190 & 1 & 4.63 & 0.75  & 380 & 200 & B1.5III             \\
  99115271 & 193911    &  ROT             &  1  &  5.560 & 15330 & 1 & 3.41 & 0.529 & 190 & 200 & B7IV/Ve shell       \\
 120967488 & 178475    &  ROT             &  1  &  5.249 & 18950 & 1 & 3.48 & 1.0   & 260 & 230 & B5/8                \\                              
 139385056 & 58978     &  BCEP            &  1  &  5.560 & 28000 & 1 & 4.41 & -     & -   & 280 & B0.5IVn(e)p         \\
 139472176 & 14850     &  ROT             &  1  &  8.400 & 13500 & 7 & 3.39 & 0.773 & 350 & 200 & B7III/IVe           \\
 140214221 & 37795     &  ROT             &  2  &  2.652 & 12200 & 2 &      & 1.841 &     & 180 & B7IVe               \\                              
 144028101 & 135734    &  MAIA            &  1  &  4.274 & 13520 & 1 &      & -     &     & 279 & B8Ve                \\                              
 148316007 & 49319     &  BCEP+ROT        &  2  &  6.625 & 24380 & 1 & 3.09 & 2.5   & 400 & 245 & B2/3IVne            \\
 148917425 & 109387    &  MAIA+ROT        &  2  &  3.890 & 14000 & 6 & 3.11 & 0.882 & 270 & 200 & B6III(n)            \\
 151300497 & 155806    &  -               &  1  &  5.530 & 34600 & 1 & 5.48 & -     &     & 115 & O7.5V((f))z(e)      \\                              
 159117671 & 112028    &  ROT+ROT         &  1  &  5.350 &  9650 & 1 & 2.14 & 1.252 & 270 & 200 & A1IIIp shell        \\
 174664153 & 61925     &  ROT             &  2  &  6.004 & 22570 & 1 & 4.02 & 1.019 & 350 & 200 & B3IV(e);            \\
 175523591 & 63215     &  ROT             &  2  &  5.870 & 17140 & 1 & 2.90 & 2.5   & 400 & 271 & B6Vnn               \\
 195744427 & 199629    &  ROT             &  1  &  3.940 &  9900 & 1 & 1.96 & -     &     & 219 & A0.5IIIn            \\   
 207176480 & 19818     &  ROT+FLARE       &  2  &  9.060 & 11710 & 1 & 1.59 & 0.298 &  90 &     & B9/A0Vne:           \\
 230981971 & 10144     &  ROT             &  1  &  0.460 & 15000 & 8 & 3.48 & 0.75  & 350 & 260 & B4V(e)              \\
 234230792 & 49330     &  BCEP+ROT        &  1  &  8.950 & 27200 & 7 & 4.16 & 1.47  & 400 & 200 & B0:nnep             \\
 245286665 & 192044    &  ROT             &  1  &  5.920 & 15330 & 1 & 3.28 & 1.039 & 330 & 280 & B7IV/V:ne shell     \\                              
 258704817 & 129954    &  -               &  1  &  5.880 & 24380 & 1 & 3.99 & -     & -   & 180 & B2.5V               \\
 259449942 & 60855     &  ROT             &  1  &  5.700 & 20760 & 1 & 3.91 & 1.1   & 390 & 230 & B4III:n shell       \\
 260640910 & 46860     &  SPB+ROT         & 12  &  5.707 & 13520 & 1 & 2.70 & 1.392 & 290 & 200 & B8III               \\
 270219259 & 209014    &  MAIA            &  1  &  5.620 & 12200 & 1 & 2.96 & -     & -   & 350 & B8III shell         \\
 277103567 & 37935     &  ROT             &  3  &  6.281 &  9940 & 5 & 2.30 & 1.497 & 360 & 209 & B9.5V               \\
 279430029 & 53048     &  ROT             & 13  &  7.920 & 18950 & 1 & 3.64 & 1.784 & 550 &     & B5/7Vn(e:)          \\
 281741629 & CD-56~152 &  -               &  1  & 10.180 & 19000 & 4 & 4.59 & -     & -   & 180 & sdB?/Be?            \\
 296969980 & 131492    &  SPB             &  1  &  5.110 & 20000 & 2 & 3.92 & -     & -   & 100 & B2IV/V              \\
 302962039 & 78764     &  -               &  4  &  4.654 & 19000 & 2 & 4.19 &  -    &     & 140 & B2IV:n(e) He-s      \\                              
 308748912 & 68423     &  -               &  6  &  6.313 & 12100 & 2 & 2.63 & -     & -   &  26 & B7IVek;             \\
 334776134 & 91120     &  ROT             &  1  &  5.580 & 11453 &10 & 2.41 & 1.3   & 270 & 250 & B9IV/V shell        \\                              
 341040849 & 64831     &  ROT             &  4  &  7.830 & 13520 & 1 & 2.61 & 1.406 & 260 &     & B8Vn(e)             \\                              
 355653322 & 224686    &  ROT             &  1  &  4.470 & 10500 & 2 & 2.39 & 1.266 & 300 & 275 & B9IIIn              \\
 358467471 & 65663     &  -               &  5  &  6.741 & 13520 & 1 & 3.48 & -     &     & 120 & B8IIIe              \\                              
 363748801 & 149671    &  ROT             &  1  &  5.882 & 13520 & 1 & 2.82 & 1.555 & 370 & 230 & B8:V:               \\
 364398342 & 66194     &  ROT             &  7  &  5.810 & 20632 & 3 & 3.76 & 1.25  & 380 & 200 & B2IVn(e)p(Si)       \\
 405520863 & 110335    &  ROT             &  1  &  4.940 & 17140 & 1 & 3.52 & 1.298 & 430 & 244 & B6IVe               \\
 408382023 & 83953     &  MAIA+ROT        &  1  &  4.760 & 15000 & 2 & 2.93 & 1.55  & 340 & 260 & B6V(e)              \\                              
 409358619 & 124367    &  BCEP+ROT        &  1  &  5.070 & 19650 & 2 & 2.97 & 2.768 & 370 & 280 & B5:Vnne             \\
 423528378 & 107348    &  ROT             &  1  &  5.210 & 12830 & 2 & 2.35 & 2.306 & 350 & 250 & B8Vn                \\
 425224332 & 58715     &  ROT             &  1  &  2.890 & 12560 &11 & 2.48 & 1.61  & 300 & 260 & B8IV/Ve shell       \\
 427395049 & 37041     &  -               &  1  &  6.390 & 27500 & 2 & 4.00 & -     & -   & 130 & O9.5Vpe             \\
 439397894 & 225132    &  ROT             &  1  &  4.543 &  9900 & 1 & 2.18 & 1.441 & 300 & 211 & A0:IV               \\
 443616529 & 98058     &  DSCT            &  1  &  4.467 &  8200 & 2 & 1.69 & -     &     & 230 & A7IV shell          \\                              
 452163402 & 100673    &  ROT             &  1  &  4.614 & 11710 & 1 & 2.60 & 1.364 & 330 & 125 & B9V(e?)             \\
 463103957 & 88661     &  ROT             &  2  &  5.750 & 25350 & 9 & 4.10 & 1.36  & 400 & 220 & B2IVnep             \\                              
 469421586 & 195554    &  MAIA+ROT        &  1  &  5.889 & 13520 & 1 & 2.85 & 1.66  & 400 & 240 & B8IV/V              \\                              
\hline
\multicolumn{12}{l}{References to $T_{\rm eff}$:}\\
\multicolumn{12}{l}{1 - MK Type \citep{Pecaut2013}; 2 - \citet{Arcos2018}; 3 -
\citet{Silaj2014}; 4 - \citet{Silva2011}; }\\
\multicolumn{12}{l}{5 - \citet{Balona1994a}; 6 - \citet{Saad2004}; 7 -
\citet{Levenhagen2006}; 8 - \citet{DeSouza2014};}\\
\multicolumn{12}{l}{9 - \citet{Zorec2016}; 10 - \citet{Shokry2018};
11 - \citet{Harmanec2019}.}\\
\hline                        
\end{tabular}
}
\end{table*}

In this paper we present photometric {\it TESS} observations of classical Be 
stars observed in Sectors 1--15.  Observations of the most interesting stars and 
stars with the longest time series are presented in the main body of the text.
Discussion of the remaining stars is deferred to an Appendix.

Our aim is to investigate the morphology of the light curves.  Where it is 
possible, we attribute the frequencies or frequency groups to rotational 
modulation.  Our aim is to test this idea by comparing the projected rotational
velocity, $v\sin i$, with the equatorial rotational velocity, $v$, derived
from the photometric frequency and the stellar radius.  For this 
purpose we obtain stellar radius estimates using the luminosity derived from 
the {\it GAIA} DR2 parallax \citep{Gaia2016,Gaia2018} and the effective 
temperature.  If this test passes, it paves the way for a new hypothesis in
which active regions may be the source of mass loss.  An idea, first suggested 
by \citet{Balona2003b}, may serve as a framework for such an hypothesis.

\begin{figure}
\begin{center}
\includegraphics{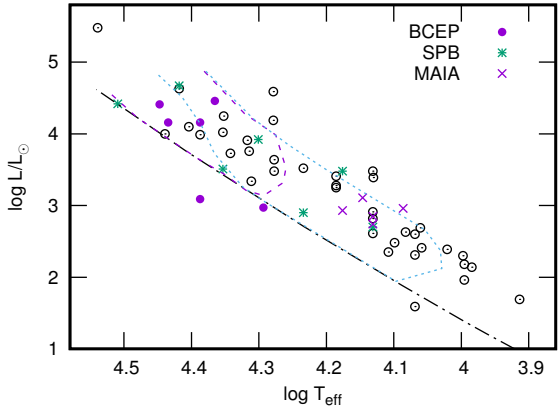}
\caption{The H-R diagram showing the Be stars observed by {\it TESS}. 
Be stars which are also $\beta$~Cep (BCEP), SPB or Maia pulsators are
indicated.  Also shown is the theoretical zero-age main sequence (solid 
line) and the instability regions of the $\beta$~Cep and SPB pulsating stars.}
\label{hrdiag}
\end{center}
\end{figure}

\section{The {\it TESS} data}

The {\it TESS} satellite obtains precise wide-band photometry for thousands
of stars with two-minute cadence in a given sector of the sky.  There are 26 
partially overlapping sectors and each sector is observed for approximately 
one month.  The light curves are obtained using pre-search data conditioning 
(PDC) which corrects for time-correlated instrumental signatures in the light 
curves \citep{Jenkins2016}.  Each {\it TESS} pixel is 21\,arcsec in size which 
is similar to the typical aperture size used in ground-based photoelectric 
photometry.  The chance of contamination by a star of similar brightness is 
not negligible. 

The stars selected for analysis were chosen from the {\tt BESS} database of
classical Be stars \citep{Neiner2011}.  These are listed in Table\,\ref{data}.
The effective temperatures, $T_{\rm eff}$, of Be stars are poorly known 
because the line blanketing caused by circumstellar material leads to
unreliable estimates from multicolour photometry.  For this reason the
value of $T_{\rm eff}$ was derived from spectroscopic modelling whenever
possible.  Failing this, it is estimated from the spectral type and luminosity 
class using the \citet{Pecaut2013} calibration. The error in $T_{\rm eff}$, as 
estimated from the dispersion in spectral types, is about 1000\,K.

Stellar luminosities were derived from {\it GAIA} DR2 parallaxes 
\citep{Gaia2016,Gaia2018}.  The bolometric correction was obtained 
from $T_{\rm eff}$ using the \citet{Pecaut2013} calibration.  The 
reddening correction was derived from a three-dimensional reddening map  by 
\citet{Gontcharov2017}.  The formal error in luminosity, $\log L/L_\odot$,
as estimated from the error in the parallax, bolometric correction and
extinction, is about 0.05\,dex, but is likely to be larger.  The theoretical 
\mbox{H--R} diagram is shown in Fig.\,\ref{hrdiag}.

The projected rotational velocities in Table\,\ref{data} are mostly from
the catalogue of \citet{Glebocki2005} supplemented by more recent 
measurements when available.  The typical error in $v\sin i$ for B stars can 
be estimated from the catalogue.  The error increases with $v\sin i$ and
ranges up to 60\,km\,s$^{-1}$.  A representative value of
$\sigma_{v\sin i} = 30$\,km\,s$^{-1}$ is assumed.  From the error in
$\log{L/L_\odot}$ and $T_{\rm eff}$ it is easy to calculate the error in
the derived stellar radius and hence the equatorial rotational velocity,
$v$, deduced from the photometric rotational frequency, $\nu_{\rm rot}$.  This 
error depends almost entirely on the error in $T_{\rm eff}$.  The contribution 
from the luminosity error is small while the contribution from the error in 
$\nu_{\rm rot}$ is negligible.  The typical value for the error in the derived 
equatorial rotational velocity is $\sigma_{v} \approx 40$\,km\,s$^{-1}$. 

\subsection{ Photometric Data and Analysis}

To detect periodic stellar signals, the Lomb-Scargle \citep{Lomb1976,
Scargle1982} technique is used.  However, the power spectrum does not 
provide information related to changes of frequency and amplitude with time. 
Such information may be very important in diagnosing the source of the
variation and are best represented by a time-frequency diagram 
\citep{Boashash2015}.

In obtaining the time-frequency diagram, the Lomb-Scargle periodogram was
calculated in a fixed time window which was chosen to be five days as a 
compromise between good signal-to-noise ratio and reasonable frequency 
resolution. The window is shifted with a time step of 0.5\,d.  At each time 
step, the periodogram (amplitude vs frequency) is plotted.  In the resulting 
time-frequency diagram, the amplitude is coded by color or grey level.

\section{Variability classification}

In the {\it General Catalogue of Variable Stars} (GCVS, \citealt{Samus2017}), 
the variability type BE is used for Be stars with mild to moderate light 
outbursts.  The GCAS ($\gamma$~Cas) type is used for eruptive early Be stars
where outbursts may exceed 1\,mag. These variations refer to the effects of 
the outburst itself and to circumstellar material on the light curve.  There 
is no GCVS classification for hot rotational variables other than those with 
chemical peculiarities.  For this reason, the ROT class was introduced. For a 
brief discussion on classification of variable stars see \citet{Balona2019a}.  

While starspots may be present in Be stars, and may be detected in the light
curve, circumstellar material is an important contribution to the light 
variation.  The tentative framework that we adopt in describing the 
quasi-periodic light variations in Be stars is that these are a result of 
co-rotating gas clouds ejected by active regions associated with starspots, 
rather than the starspots themselves. They therefore give rise to incoherent
light variations with a quasi-period close to the rotation period of the
star.  In the periodogram this manifests as groups of peaks or broad humps
of power at the approximate rotational frequency and its harmonic. 

The Be stars lie in either, or both, the $\beta$~Cep or SPB instability 
strips and it is likely that pulsations may be present.  Some stars have high 
frequencies, but are too cool to be classified as $\beta$~Cep and too hot for 
$\delta$~Scuti.  These are classified as Maia variables \citep{Balona2016c}.  
It is more difficult to distinguish SPB pulsations from rotation modulation as 
the frequency range overlaps.  The criterion used in that if harmonics are 
present then rotation cannot be excluded and the ROT class is applied.  From 
the fundamental photometric frequency, the equatorial rotational velocity is 
estimated.  This will be compared with $v\sin i$ as a consistency test.

\section{Results}

There are several stars where {\it TESS} observations cover many sectors. 
These are of particular interest as it allows exploration of how the star
behaves over a period of many months.  Certain well-observed stars are
also of interest even if the observations are only from one sector.  These
stars are discussed below.  The remaining stars, nearly all of which were 
observed in only one sector, are discussed in the Appendix.   

\begin{figure}
\begin{center}
\includegraphics{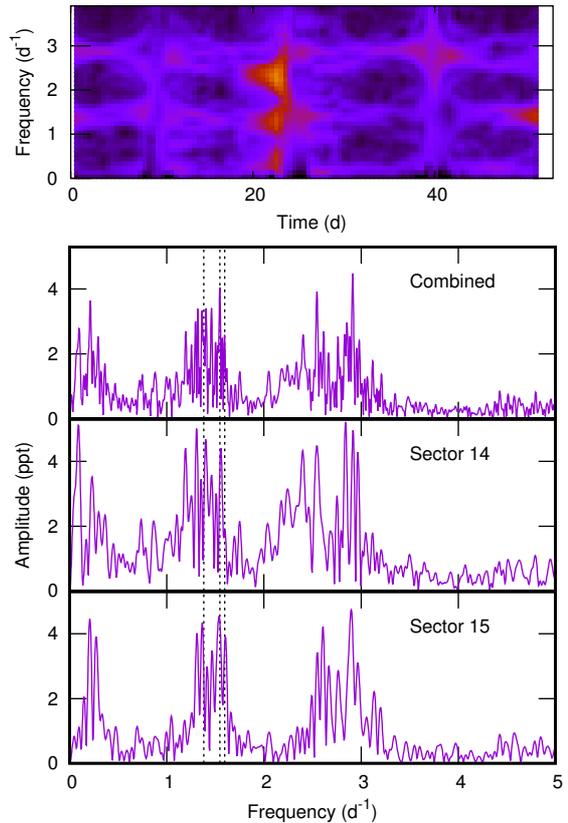}
\caption{Time-frequency diagram and periodograms of the combined data and of 
individual observing sectors of TIC\,42360166 (28\,Cyg).  The vertical dotted 
lines represent the frequencies 1.381, 1.545 and 1.597\,d$^{-1}$ mentioned
by \citet{Baade2018b}.}
\label{28Cyg}
\end{center}
\end{figure}

\subsection{TIC\,42360166, 28\,Cyg}

Spectroscopic observations of 28\,Cyg have generally confirmed the presence
of a group of frequencies at around 1.56\,d$^{-1}$ \citep{Peters1988, 
Pavlovski1990, Hahula1994} and also around 1.4\,d$^{-1}$ \citep{Spear1981, 
Bossi1993}. \citet{Tubbesing2000} confirmed the 1.56\,d$^{-1}$ frequency and 
found another group at about 1.60\,d$^{-1}$.  28\,Cyg was extensively studied 
by \citet{Baade2018b} using photometric data from the {\it BRITE} and 
{\it SMEI} satellites. The variability is clustered into three frequency groups with 
approximate ranges 0.1--0.5, 1.0--1.7 and 2.2--3.0\,d$^{-1}$. The peaks of 
highest amplitude occur at 1.381, 1.545 and 1.597\,d$^{-1}$.

\citet{Baade2018b} note that stochastic variability is an important
contribution to the light curve.  However, they stress that coherent 
variations certainly occur.  They point out the clear signature of low-order 
g-mode pulsations in high-resolution spectra \citep{Rivinius2003}, 
demonstrating coherent large-scale structures in the photosphere.  These 
observations were, necessarily, of relatively short duration so it is
not entirely clear that coherence is well established.

Periodograms of the {\it TESS} data are shown in Fig.\,\ref{28Cyg}.  These
look very different from the {\it BRITE} periodograms in that the frequency
spectrum is much denser.  However, the three groups are still present.  The 
three coherent peaks of largest amplitude mentioned by \citet{Baade2018b} do 
not seem to correspond with peaks in the {\it TESS} data. Either the amplitudes
are lower or the frequencies are not quite as coherent as expected.  Note that 
peaks in sector 14 look quite different from those in sector 15 and that the 
time-frequency diagram suggests somewhat erratic behaviour.

Because one frequency group is roughly twice the frequency of the other, we
have assumed that the fundamental group at around 1.5\,d$^{-1}$ reflects the 
approximate rotational frequency of the star.  The frequency group at
 0.1--0.5\,d$^{-1}$ can be attributed to circumstellar material.

\begin{table}
\caption{Extracted frequencies, $\nu$ (d$^{-1}$), and amplitudes, $A$ (ppt).
The figures in brackets denote the error in the last digit.}
\label{extra}
\resizebox{\columnwidth}{!}{
\begin{tabular}{rrrrrr}
\hline
\multicolumn{1}{c}{$\nu$} & 
\multicolumn{1}{c}{$A$}   &
\multicolumn{1}{c}{$\nu$} & 
\multicolumn{1}{c}{$A$}   &
\multicolumn{1}{c}{$\nu$} &
\multicolumn{1}{c}{$A$}   \\
\hline
\multicolumn{4}{l}{TIC\,65803653 (27\,CMa):}\\
   0.7896(4) & 2.09(1) &  2.6793(4) & 1.54(1) & 13.1131(9) & 0.10(1) \\
   1.2608(4) & 2.06(1) &  5.1578(6) & 0.35(1) & 13.5096(5) & 0.54(1) \\   
   1.3472(4) & 1.68(1) &  5.9020(7) & 0.31(1) & 14.7340(9) & 0.11(1) \\   
   1.4060(4) & 1.26(1) & 10.8837(4) & 1.08(1) & 24.3903(9) & 0.09(1) \\   
   2.6240(4) & 1.88(1) &            &         &            &         \\   
\hline                        
\multicolumn{4}{l}{TIC\,234230792 (HD\,49330):}\\
   0.8601(7) & 1.56(3) &  2.9164(4) & 3.34(3) &  5.8908(9) & 0.83(3) \\
   2.4565(8) & 1.31(3) &  4.2518(7) & 1.48(3) & 11.8657(9) & 0.77(3) \\
   2.6044(7) & 1.66(3) &  4.3771(9) & 0.85(3) & 16.1122(9) & 0.27(3) \\
   2.7154(7) & 2.39(3) &            &         &                      \\
\hline 
\end{tabular}
}
\end{table}

\begin{figure}
\begin{center}
\includegraphics{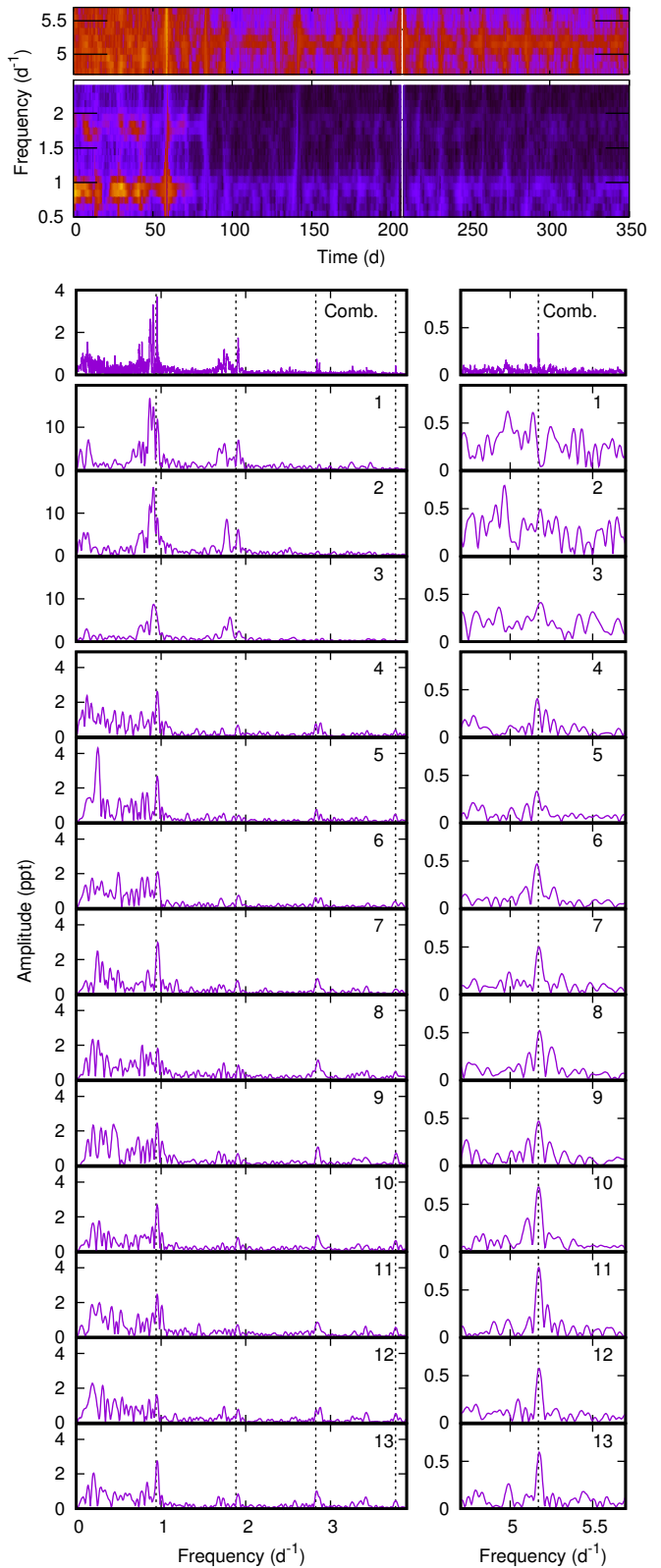}
\caption{The time-frequency diagram and periodograms of the combined data and 
of individual observing sectors (labeled) of TIC\,55295028. Note the changes
in amplitude scale in the periodograms.  The vertical dotted lines represent 
the frequency $\nu = 0.94$\,d$^{-1}$ and its three harmonics on the left.  The 
right hand panels shows the peak $\nu=5.17141$\,d$^{-1}$.}
\label{055295028}
\end{center}
\end{figure}

\subsection{TIC\,55295028, HD\,33599}

\citet{Bernhard2018} detected a frequency of 0.9051\,d$^{-1}$ from ground-based
photometry. \citet{Balona2019a} did not detect a period from {\it TESS}
sector 1--2.

The time-frequency diagram (Fig\,\ref{055295028}) shows two groups with 
fundamental frequency of around $0.94$\,d$^{-1}$ which we adopt as the
rotational frequency.  Three harmonics are visible as indicated by the dotted 
lines. There is a sudden decline in amplitude in all groups between sector 3 
and 4. 

There is also a low-amplitude peak at 5.171\,d$^{-1}$ which is first visible 
in sector 4 and subsequent sectors.  The line is sharp and the frequency 
constant (right panel in Fig.\,\ref{055295028}).  It does not seem to be 
harmonically related to the broad peak groups and is most probably due to 
pulsation.  The star may thus be classified as a $\beta$~Cep variable. It is 
possible that the $\beta$~Cep pulsation retains constant amplitude, but the 
increased background noise in sectors 1--3 renders it less visible.

\begin{figure}
\begin{center}
\includegraphics{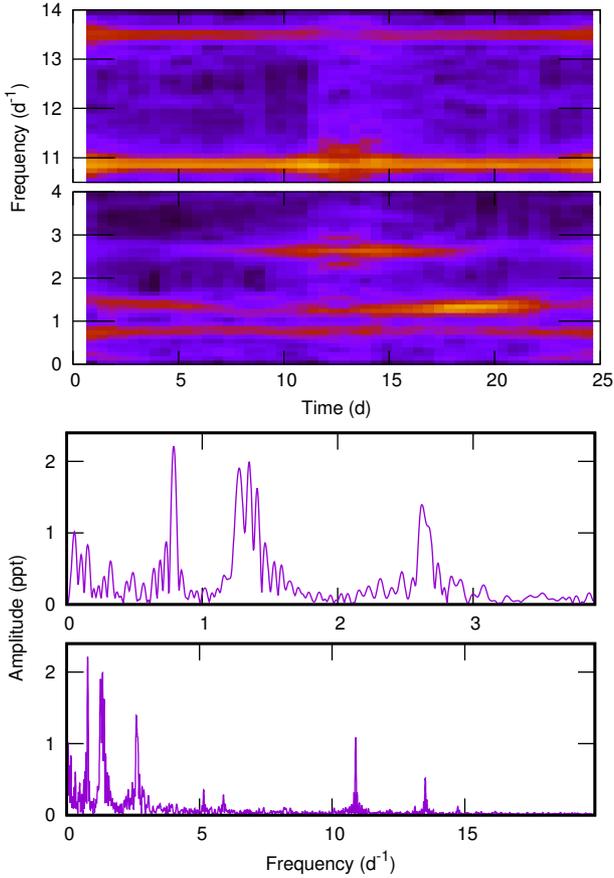}
\caption{Time-frequency diagrams and periodograms of TIC\,65803653
(27~CMa) showing high frequencies as well as variable-amplitude peaks.}
\label{27CMa}
\end{center}
\end{figure}

\subsection{TIC\,65803653, 27~CMa}

27\,CMa is a very close optical multiple system. From ground-based photometry, 
\citet{Balona1991a} found a single frequency, 0.7925\,d$^{-1}$, in 1986 
and 1987.  When the star was next observed in 1990, in addition to the quoted 
frequency, a new peak at 10.8914\,d$^{-1}$ made its appearance.  The high 
frequency, characteristic of a $\beta$~Cep star, was still present in 1991.

The appearance of $\beta$~Cep pulsations may be related to the changes in the
spectrum reported by \citet{Bhattacharyya1989a} and \citet{Bhattacharyya1989b}.
Prior to the appearance of $\beta$~Cep pulsations, observations showed typical 
double-peaked H$\alpha$ emission.  In early 1989 the H$\alpha$ and 
HeI\,5876 lines displayed P-Cyg profiles.  Later that year, the appearance of 
the spectrum in the region of H$\alpha$, Si\,6347, Si\,6371 and HeI\,5876 
suggested that the star had entered into a shell phase.

Since self-driven pulsations are not known to appear and disappear in non-Be 
stars, it is perhaps possible that the pulsations were simply masked from
view prior to the development of the shell phase.  Once the shell phase was
established, the obscuration was removed and the $\beta$~Cep pulsations
became visible.

The {\it TESS} periodogram (Fig.\,\ref{27CMa}) shows multiple peaks in which 
$\beta$~Cep pulsations are visible.  A list of extracted frequencies is shown 
in Table\,\ref{extra}.  The peak at 10.8837\,d$^{-1}$ corresponds to that
seen by \citet{Balona1991a}.  The time-frequency diagram indicates variability 
in both amplitude and frequency in some of the low-frequency peaks.  By
contrast, the frequency and amplitude of the $\beta$~Cep pulsations seem to
be stable.  The peak at 2.6\,d$^{-1}$ seems to be an harmonic of the broad peak
at around $\nu = 1.3$\,d$^{-1}$ which is assumed to be the rotational frequency.
The nature of the 0.7896\,d$^{-1}$ frequency, which is the one observed by 
\citet{Balona1991a}, is not clear.  It appears to be stable in both
frequency and amplitude.

\begin{figure}
\begin{center}
\includegraphics{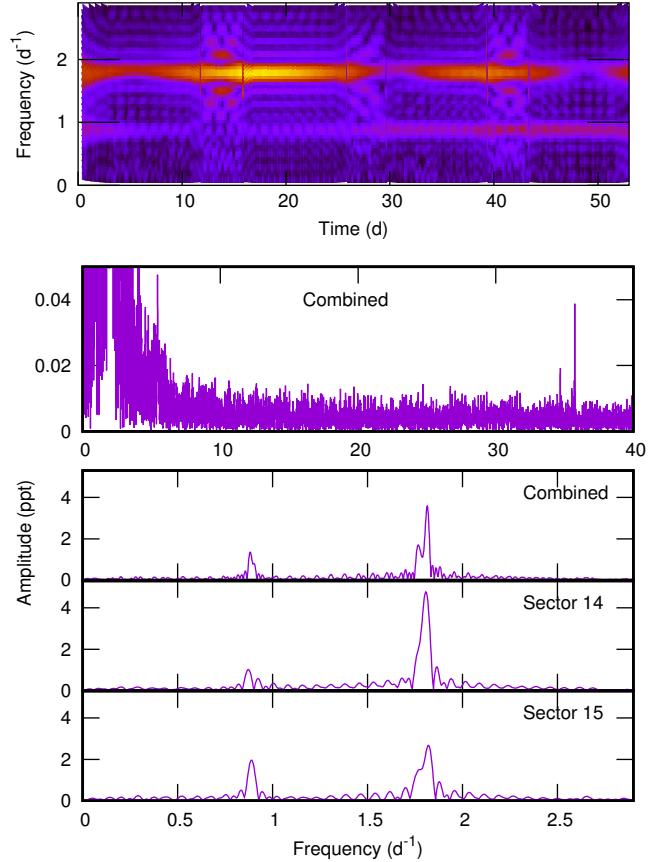}
\caption{Time-frequency diagram and periodograms of TIC\,148917425 ($\kappa$~Dra) 
showing high frequencies (top panel) as well as variable-amplitude rotational
peaks.} 
\label{KappaDra}
\end{center}
\end{figure}

\subsection{TIC\,148917425, $\kappa$~Dra}

\citet{Baker1920} found $\kappa$~Dra to be a spectroscopic binary with a 
period of 8.986\,d (0.111\,d$^{-1}$).  Later \citet{Hill1926} confirmed the 
binary nature but with a period of 0.89038\,d (1.123\,d$^{-1}$) which is the 
1-d alias of the period found by Baler.  \citet{Juza1991} obtained further 
spectroscopic observations and confirmed the presence of the 0.89-d period 
which is transient in nature and suggested that it might be the rotation 
period.  In addition, they found the star to be a binary with a period of 
61.55\,d.

The {\it TESS} periodograms (Fig,\,\ref{KappaDra}) show two broad peaks at a 
frequency of $\nu \approx 0.89$\,d$^{-1}$ and its harmonic.  The 0.89\,d$^{-1}$
peak is the 1-d alias of the 1.123\,d$^{-1}$ frequency found by 
\citet{Juza1991}.  The time-frequency diagram suggests that their amplitudes 
vary, in agreement with the transient character reported by
\citet{Juza1991}.

In addition, the {\it TESS} periodogram of the combined data shows a pair of 
high-frequency peaks at 34.696 and 35.744\,d$^{-1}$.  This would suggest that 
$\kappa$~Dra is a Maia star \citep{Balona2016c}, since it is too cool for a 
$\beta$~Cep variable.

\subsection{TIC\,207176480, HIP\,14595} 

\citet{Houck1975} classified  (HD\,19818) as B9/A0Vne: with weak Balmer 
emission.  It is an X-ray source \citep{Naze2018}.  From ground-based 
{\it KELT} photometry, \citet{Oelkers2018} obtained a frequency of 
0.7031\,d$^{-1}$.

\begin{figure}
\begin{center}
\includegraphics{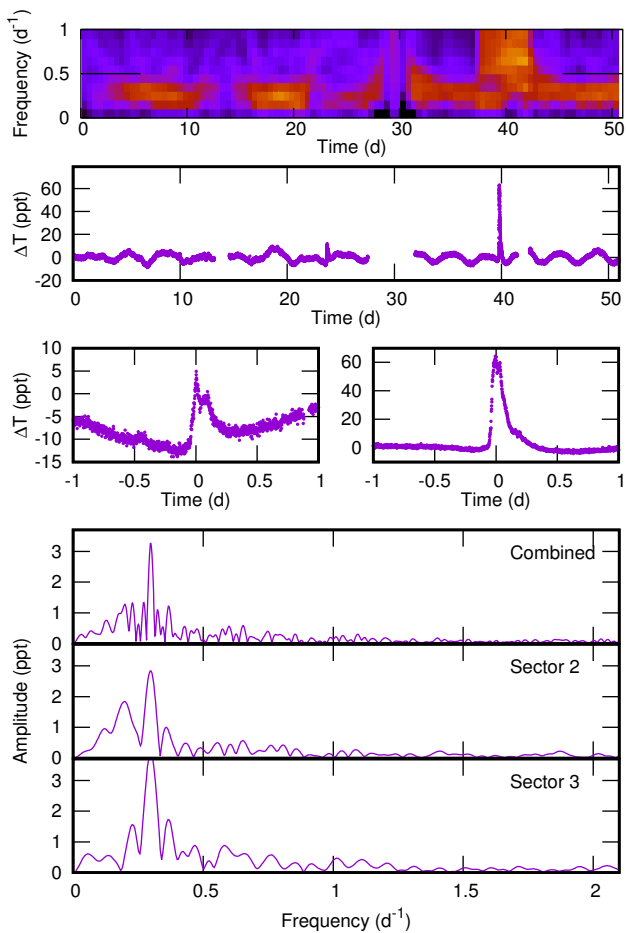}
\caption{The top panel is time-frequency diagram of TIC\,207176480
(HD\,19818) showing amplitude changes in the 0.2977\,d$^{-1}$ frequency. The 
next three panels show the light curve and detail of two flares.  The bottom 
three panels show periodograms of the combined data and periodograms of data 
in two sectors.}
\label{207176480}
\end{center}
\end{figure}

\citet{Balona2019a} list the star, but no periodicity was found from {\it
TESS} sector 2 alone.  The additional sector 3 observations show two strong 
flares at BJD\,2458377.72 and BJD\,2458393.80 (Fig.\,\ref{207176480}).  These 
are probably related to the X-ray source.  On removing the flares, the 
resulting periodogram shows just a single peak at 0.2977\,d$^{-1}$ (amplitude 
2.740\,ppt). The {\it KELT} frequency is the 1-d alias of the this frequency. 
This is assumed to be the rotational frequency, but it could also be an orbital
frequency in which the secondary is involved in generating the X-rays 
and flares.  The projected rotational velocity is not known.  It would be
important to obtain more detailed spectroscopic observations of this 
interesting star.

\subsection{TIC\,230981971, $\alpha$~Eri}

Achernar (HD\,10144) is the brightest and the nearest Be star in 
the sky.  Using Earth-rotation synthesis on the {\it VLT} interferometer,
\citet{DeSouza2003} measured an oblateness of $1.56 \pm 0.05$.  The star
has a close companion, most likely an A1V--A3V star, with an orbital period of 
about 15\,yr \citep{Kervella2008}.  

The first indication of short-period variations was a photometric/spectroscopic 
study by \citet{Balona1987c} who found a frequency of 0.79\,d$^{-1}$.  From 
line profile observations,\citet{Rivinius2003} obtained 0.77\,d$^{-1}$.  
\citet{Goss2011} analysed photometry from the {\it SMEI} instrument in yearly 
50-d segments over a 5-yr period.  They found frequencies of 0.725, 0.775 and 
one of much lower amplitude at 0.689\,d$^{-1}$, all with variable amplitudes.  
The 0.775\,d$^{-1}$ frequency appears to be coherent over the whole period,
while the 0.725\,d$^{-1}$ frequency did not exhibit coherence.
\citet{Balona2019a} did not detect a period from {\it TESS} sectors 1--2.

\begin{figure}
\begin{center}
\includegraphics{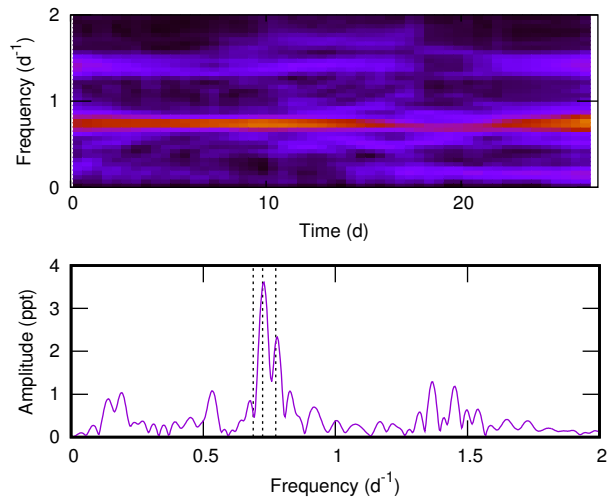}
\caption{The top panel shows the time-frequency diagram of TIC\,230981971
($\alpha$~Eri) from {\it TESS} sector 2 data.  The bottom panel is the
corresponding periodogram.  The dotted lines are the frequencies 0.689,
0.725 and 0.775\,d$^{-1}$ found by \citet{Goss2011}.}
\label{AlpEriFrq}
\end{center}
\end{figure}

The periodogram from one sector of {\it TESS} observations show two peak
groupings at around 0.73 and 1.46\,d$^{-1}$.  Each group shows multiple 
closely-spaced peaks (Fig.\,\ref{AlpEriFrq}). The {\it TESS} periodogram is 
very similar to the one obtained by \citet{Goss2011} and the frequencies at
0.725 and 0.775\,d$^{-1}$ seem to be still present (dotted lines in the figure).
Judging from the time-frequency diagram, amplitude variations are 
present in both frequency groups.  The peaks at 1.45\,d$^{-1}$ also appear 
to change frequency.

\citet{DeSouza2014} derive the stellar parameters from interferometric 
observations, finding an equatorial radius $R_{\rm eq} = 9.16 \pm 0.20
R_\odot$.  They adopt 0.689\,d$^{-1}$ as the rotational frequency, which is
the lowest-amplitude frequency found by \citet{Goss2011}, because it fits
better with their calculations. This leads to $v/v_c \approx 0.81$. 
If we assume that 0.73\,d$^{-1}$ represents the rotational frequency (with
the 1.46\,d$^{-1}$ the first harmonic) and a radius of 9.16\,$R_\odot$, an 
equatorial rotational velocity of 340\,km\,s$^{-1}$ and $v/v_c \approx 0.74$ 
is obtained. The projected rotational velocity is $v\sin i = 260 
\pm 15$\,km\,s$^{-1}$ \citep{DeSouza2014}.

The actual $v\sin i$ of Achernar seems to significantly vary in time, as 
shown by \citet{Rivinius2013b}.  The reported values span the range
223--290\,km\,s$^{-1}$.  The value appears to correlate with the disk
emission, being higher at when emission is strongest. It is interpreted as a 
variation of the equatorial rotation rate of photospheric layers and further
elaborated by \citet{Rivinius2016c}. 

\begin{figure}
\begin{center}
\includegraphics{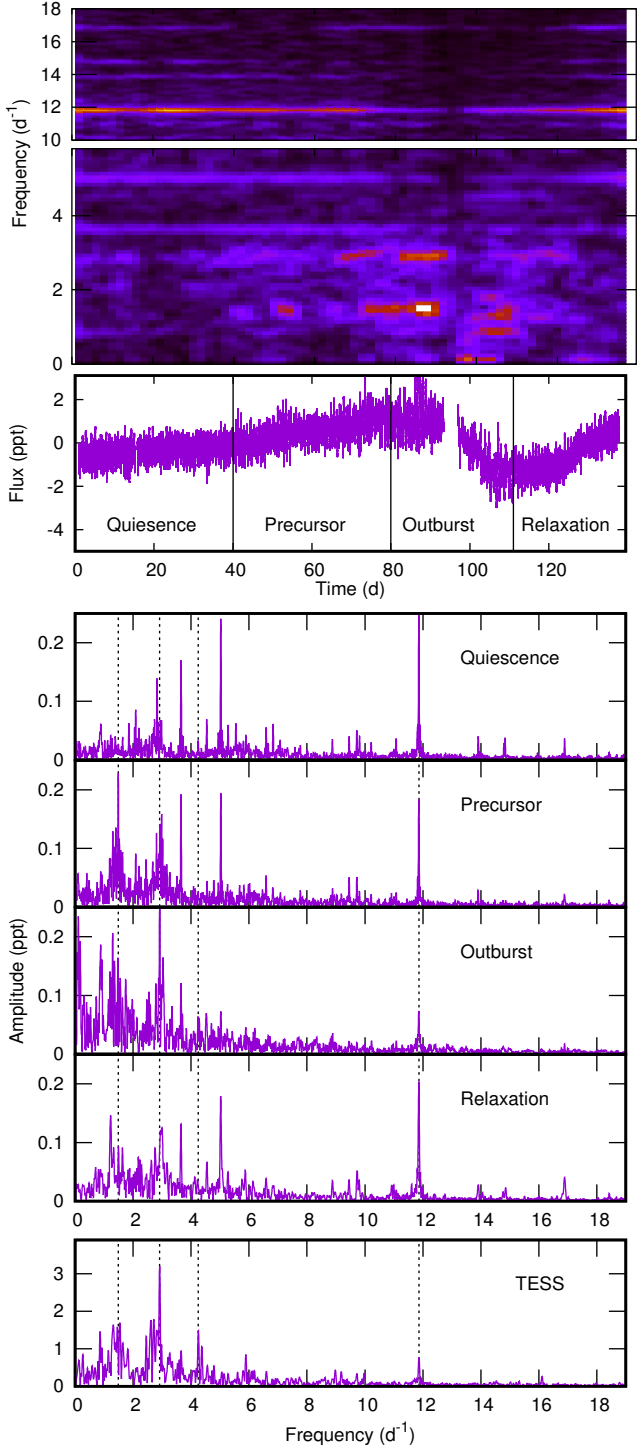}
\caption{The top panels are time-frequency diagrams for {\it CoRoT}
observations of TIC\,234230792 (HD~49330).  The next panel is the {\it CoRot} 
light curve showing four phases of the mini-outburst.  Below this panel, the 
{\it CoRoT} periodograms corresponding to each outburst phase are shown.
The bottom panel is the {\it TESS} sector 6 periodogram.  The dotted lines
are at frequencies 1.4934, 2.9164, 4.2518 and 11.8657\,d$^{-1}$.}
\label{HD49330frq}
\end{center}
\end{figure}

\begin{figure}
\begin{center}
\includegraphics{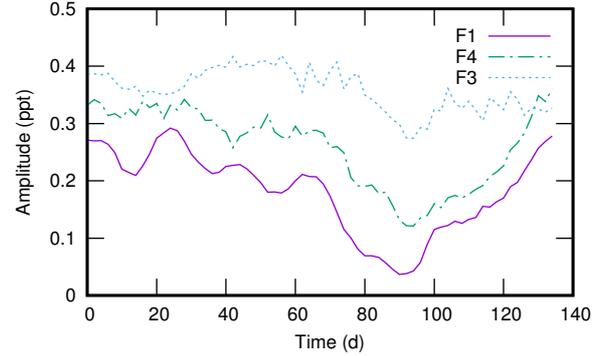}
\caption{Variation of the amplitudes of the $\beta$~Cep modes with
frequencies $F1 = 11.8639$, $F3 = 3.6594$ and $F4 = 5.0260$\,d$^{-1}$ as a
function of time from the {\it CoRoT} data of HD\,49330.  The amplitudes
have been arbitrarily displaced to reduce confusion.}
\label{HD49330amp}
\end{center}
\end{figure}

\subsection{TIC\,234230792, HD\,49330}

HD\,49330 was observed by the {\it CoRoT} mission for nearly 137\,d during
2007--2008, as described by \citet{Huat2009}.  During this time the star 
underwent a mini outburst where the brightness increased by 0.03\,mag over the 
course of a 20-d period (Fig.\,\ref{HD49330frq}). Prior to, during, and after 
the outburst, \citet{Huat2009} identify four phases as shown in 
Fig.\,\ref{HD49330frq}.

The periodogram of the {\it TESS} data, also shown in Fig\,\ref{HD49330frq}, 
closely resembles the outburst phase. Extracted frequencies are listed in 
Table\,\ref{extra}.

During the precursor and outburst phases two groups of peaks around 1.47 and 
2.94\,d$^{-1}$ gradually emerge. In addition, a more concentrated group 
around 0.87 and 1.28\,d$^{-1}$ also makes its appearance.  The amplitudes
of the high-frequency $\beta$~Cep pulsations at 11.86 and 5.03\,d$^{-1}$ 
decrease remarkably during the outburst phase.

\begin{figure}
\begin{center}
\includegraphics{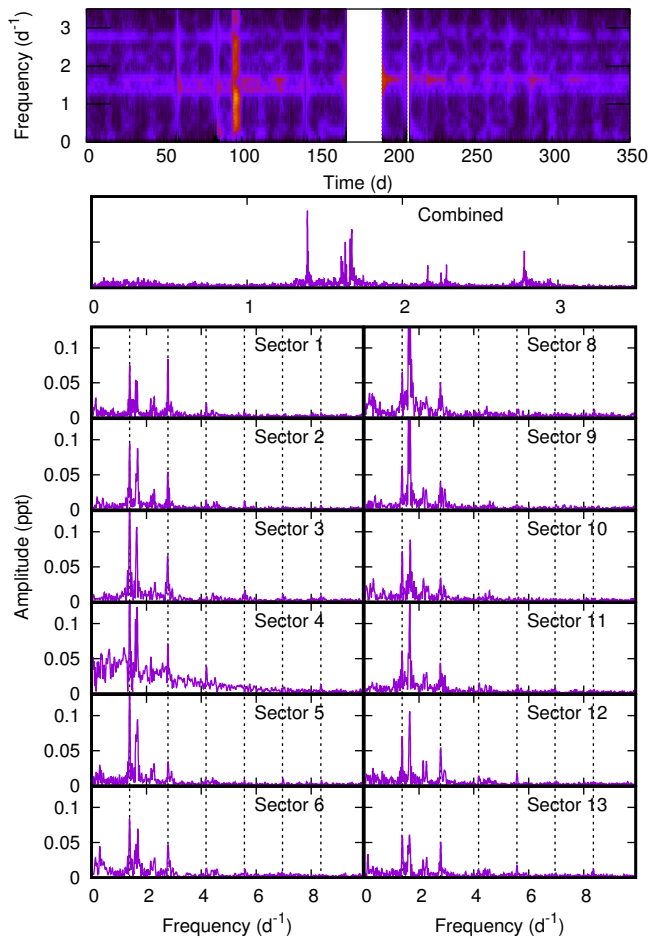}
\caption{Time-frequency diagram and periodograms of the combined data and of 
individual observing sectors of TIC\,260640910 ($\mu$~Pic).  The dotted lines 
show $nu = 1.38854(1)$\,d$^{-1}$ and 5 harmonics.}
\label{260640910}
\end{center}
\end{figure}

\citet{Huat2009} interpret the peak groups around 1.47 and 2.94\,d$^{-1}$ as
systems of closely-spaced g modes. They find a correlation between both 
amplitude changes and the presence/absence of certain frequencies of 
pulsations at the different phases of the outburst. The amplitudes of the
main frequencies (the $\beta$~Cep pulsations) decrease before and during the 
outburst and increase again after the outburst.  Also, several groups
of frequencies (g modes) appear just before the outburst, reach maximum
amplitude during the outburst and then disappear as soon as the outburst has 
finished.  As already mentioned, these are interpreted as g modes with short 
lifetimes.  They suggest that the frequency group around 0.87\,d$^{-1}$,
which is compatible with the rotational frequency, could be explained by the 
ejection of material co-rotating with the star.

The interpretation by \citet{Huat2009} is in line with the common view that
NRP is the cause of the mass loss \citep{Rivinius1998b}.  The idea is that the 
sudden large increase in pulsation amplitudes of the groups around 1.47 and 
2.94\,d$^{-1}$ is responsible for the ejection of material. The decrease in 
amplitude shortly afterwards shuts off further mass loss.  However, no
explanation has ever been given of why pulsation modes should suddenly change 
in amplitude.

\begin{figure}
\begin{center}
\includegraphics{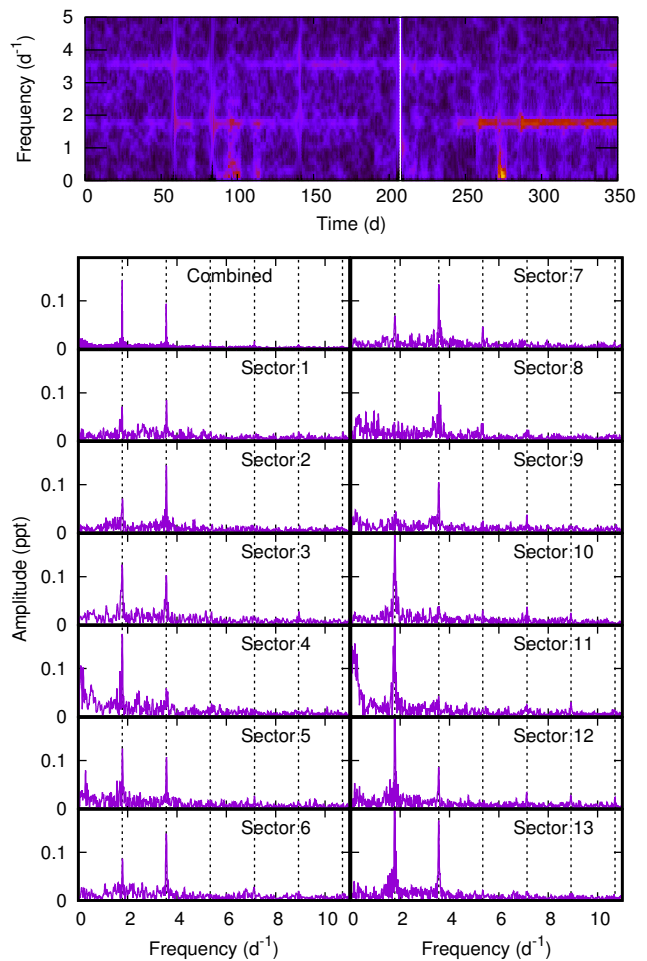}
\caption{Time-frequency diagram and periodograms of the combined data and of 
individual observing sectors of TIC\,279430029. The vertical dotted lines 
represent the frequency $\nu = 1.784$\,d$^{-1}$ and its harmonics.}
\label{279430029}
\end{center}
\end{figure}

In our view, this interpretation is not consistent with the available
information.  In the first place, it is clear that Be stars, in general, do 
not rotate very close to the critical rotation rate, as required by the
notion that pulsation acts as a trigger for mass loss.  Also, the so-called
g-mode groups around 1.47 and 2.94\,d$^{-1}$ appear to be non-coherent as is
evident from the time-frequency diagram of Fig.\,\ref{HD49330frq}.  
Note the appearance of these frequency groups compared to the smooth
appearance of the $\beta$~Cep pulsations at 3.6594, 5.0260 and
11.8639\,d$^{-1}$.  Finally,  no pulsating star so far observed shows such 
rapid increase and decrease of pulsation amplitudes.  Amplitude changes in all 
known self-excited pulsating stars occur over timescales of many months to 
years and never in just a few days.

An alternative interpretation that may be considered is that there is an 
active region on the star which is responsible for driving mass loss.  The
ejected gas clouds obscure light and results in rotational light modulation. 
We can identify the rotational frequency as approximately 1.47\,d$^{-1}$ with 
2.94\,d$^{-1}$ being the harmonic (and not an independent group of g modes).  
The rapid increase of amplitude of these two frequencies is caused by 
the build-up of ejected material co-rotating with the star.  The ejected 
material masks the photosphere, resulting in a decrease of visibility of the 
$\beta$~Cep pulsations during the outburst. As the material disperses, the 
photosphere becomes visible and the $\beta$~Cep pulsations regain their 
original amplitudes.

That the decrease in pulsation amplitude of the $\beta$~Cep modes is an
obscuration effect is demonstrated by the variation of the pulsation
amplitudes with time for the $\beta$~Cep pulsations.  The amplitudes of these 
modes vary in exactly the same way with time as can be seen in 
Fig.\,\ref{HD49330amp}.  This is extremely unlikely to occur in the NRP 
interpretation because this requires that three independent modes, which are 
bound to have different growth rates and damping rates, change amplitude by the
same fraction as a function of time.

\subsection{TIC\,260640910, $\mu$~Pic} 

The shell star $\mu$~Pic is a double star (B9IVn + A8V:p) with a separation of 
2.4\,arcsec and a magnitude difference of 3.2\,mag. \citet{Balona1992b}
found a frequency of 2.52\,d$^{-1}$ from ground-based photometry. The first
two {\it TESS} sectors were analysed by \citet{Balona2019a}, but no
periodicity was found.

The periodogram of all 12 sectors (Fig.\,\ref{260640910}) shows multiple
peaks characteristic of an SPB pulsating variable.  The main peak at 
$1.392$\,d$^{-1}$ (amplitude  0.08\,ppt) appears to be slightly
broadened or double and of variable amplitude.  This does not bear any
relationship to the frequency mentioned by \citep{Balona1992b}.  Up to 5 
harmonics are visible as shown in Fig.\,\ref{260640910}, each harmonic being 
rather broad.  This may be the rotational frequency.  A secondary peak system at 
about 1.63\,d$^{-1}$ consists of multiple peaks of variable amplitude which 
may be due to g-mode pulsations. For this reason we classify the star as 
SPB+ROT.

\begin{figure}
\begin{center}
\includegraphics{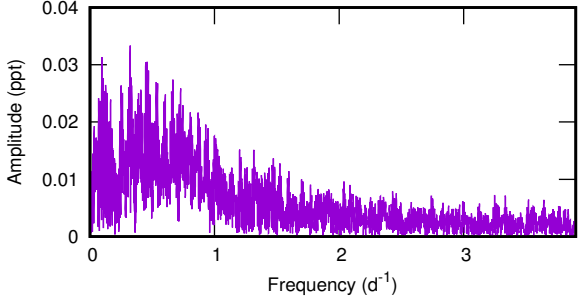}
\caption{Periodogram of combined data for
TIC\,308748912.}
\label{308748912}
\end{center}
\end{figure}

\begin{figure}
\begin{center}
\includegraphics{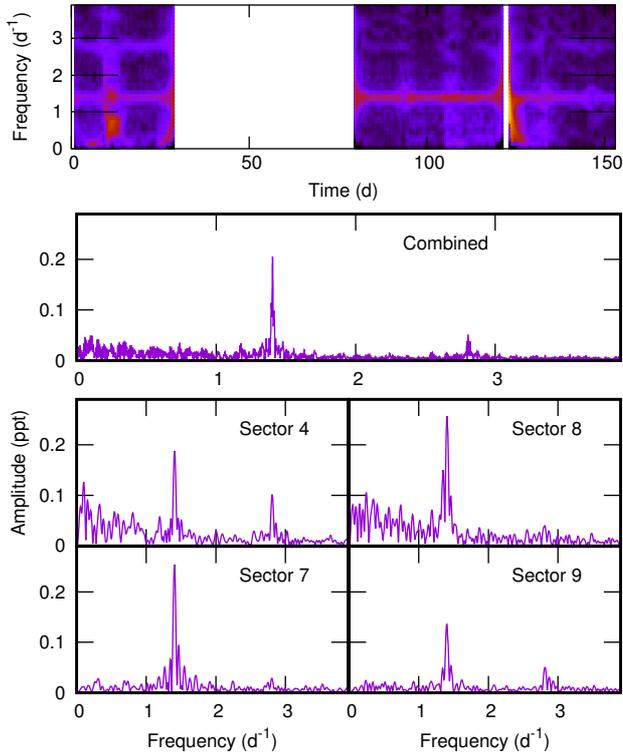}
\caption{Time-frequency diagram and periodograms of the combined data and of 
individual observing sectors of TIC\,341040849.}
\label{341040849}
\end{center}
\end{figure}

\begin{figure}
\begin{center}
\includegraphics{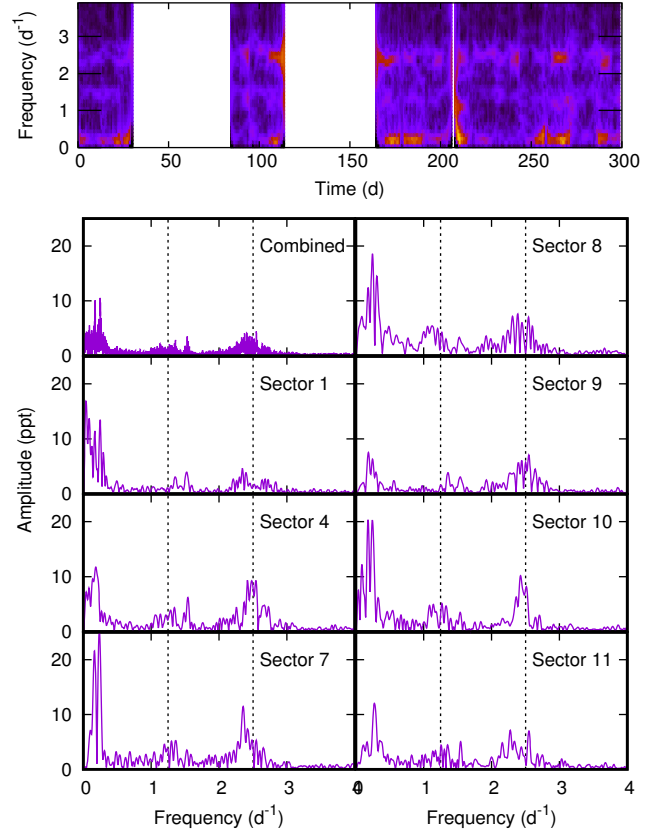}
\caption{Time-frequency diagram and periodograms of the combined data and of 
individual observing sectors of TIC\,364398342.  The vertical dotted lines 
represent the frequency $\nu = 1.25$\,d$^{-1}$ and its harmonic.}
\label{364398342}
\end{center}
\end{figure}

\subsection{TIC\,279430029, HD\,53048} 

\citet{Balona2019a} found a frequency of $1.784$\,d$^{-1}$ and its 
harmonic from first-light {\it TESS} observations.  The periodogram of 
the complete data set (13 sectors) confirms a broad or unresolved main peak 
at this frequency with amplitude of 0.14\,ppt.  Five harmonics are visible
and we assume  $1.784$\,d$^{-1}$ to be the rotational frequency.  
The amplitude of the fundamental peak and its harmonics change quite 
dramatically from sector to sector, as shown in Fig.\,\ref{279430029}.  
On occasions (sectors 8 and 9), the fundamental disappears altogether,
resulting in a double-wave light curve.  

\subsection{TIC\,308748912, HR\,3217} 

This is a poorly studied B6V or B7IVek star with narrow lines 
($v\sin i = 26$\,km$^{-1}$; \citealt{Zorec2012}).  A study of the first two
sectors from {\it TESS} by \citet{Balona2019a} showed no significant
periodicity  The addition of four further sectors does not change this 
assessment (Fig.\,\ref{308748912}).  While this star appears to be 
uninteresting, it presents an important challenge in understanding how mass 
loss can arise in a star which seems to lack any activity.

\subsection{TIC\,341040849, HIP\,38433}

This poorly-studied star is a member of NGC\,2516.  A peak at 1.406\,d$^{-1}$ 
and its harmonic are visible (Fig.\,\ref{341040849}).  Both peaks 
are rather broad and variable in amplitude.  This simple frequency spectrum
is similar to that in TIC\,55295028 (Fig.\,\ref{055295028}), $\kappa$~Dra
(Fig.\,\ref{KappaDra}) and TIC\,279430029 (Fig.\,\ref{279430029}).  The light
curve is almost sinusoidal as in $\eta$~Cen \citep{Baade2016} and $\nu$~Pup
\citep{Baade2018a}.  Light curves showing a broad or very closely spaced
multiple peaks and its harmonic are common and describe about 40\,percent of
the Be stars in our sample.  They seem to be mostly mid- or late-B stars.

\begin{figure}
\begin{center}
\includegraphics{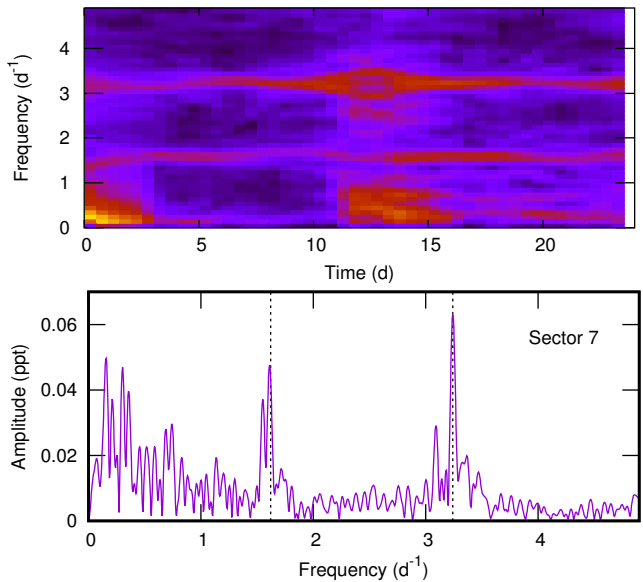}
\caption{Time-frequency diagram and periodograms of $\beta$~CMi.  The dotted
lines represent the assumed rotational frequency 1.621\,d$^{-1}$ and its
first harmonic.}
\label{betCMi}
\end{center}
\end{figure}

\subsection{TIC\,364398342, V374~Car}
 
This star is a member of the open cluster NGC\,2516 \citep{Sampedro2017} and
also an X-ray source \citep{Naze2018}.  From its position in the
\mbox{H-R} diagram it can be classified as a blue straggler \citep{Ahumada2007}
and may be a spectroscopic binary \citep{Gonzalez2000}. The star was discussed 
by \citet{Balona2019a} but no periodicity was mentioned. 

The time-frequency diagram and periodograms in Fig.\,\ref{364398342} shows two 
broad peaks at about 1.25 and 2.50\,d$^{-1}$ as well as structure at very low 
frequencies. No resolved feature can be seen. Broad peaks such as these are
found in several Be stars such as KIC\,6954726 and KIC\,11971405
\citep{Rivinius2016}.  It is possible that the broadening is simply a result
of incoherence  associated with co-rotating gas clouds close to the
photosphere and that the 1.25\,d$^{-1}$ measures the approximate rotation 
frequency of the star.

\subsection{TIC\,425224332, $\beta$~CMi}

\citet{Saio2007} analysed photometry of $\beta$~CMi from the {\it MOST} 
satellite and concluded that it is a multiperiodic non-radial pulsator.
In a re-analysis of the same data, \citet{Harmanec2019} did not find any
coherent frequency.  The dominant frequency of 3.257\,d$^{-1}$ appears to vary 
in frequency around its mean value.  They conclude that there is only one 
stable frequency of 1.621\,d$^{-1}$ which they associate with the rotational
frequency.

The {\it TESS} time-frequency diagram and periodogram is show in
Fig.\,\ref{betCMi}.  The peak at around 1.61\,d$^{-1}$ is double, but this 
is due to slight frequency variations as can be seen by close inspection of 
the time-frequency diagram.  The peak around 3.35\,d$^{-1}$ is equally
variable in frequency and amplitude. There are indications of the 4th and 6th  
harmonics as well.  Most likely, this is another case of a star in which the
rotational frequency, $1.621$\,d$^{-1}$, and its harmonic is present. 

During the {\it MOST} observations, this frequency had very low amplitude
and the light curve was dominated by the first harmonic, giving a double-wave 
variation.  At that time, the first harmonic could have mistakenly taken as
the rotational frequency.  In the {\it TESS} data, the light curve has unequal 
minima so that the true rotation frequency is revealed.

\section{Discussion}

In the Appendix, periodograms of the remaining {\it TESS} Be stars are
presented.  Out of the 57 stars in our sample, 25 have a simple periodogram 
consisting of a peak and its harmonic, or in some stars, several harmonics.  
The peak may be somewhat broadened or with fine structure. In a further 18 
stars, the peaks cluster in groups which have an harmonic relationship.  The 
peak frequencies are different from sector to sector.  Often the peaks are 
broad.  All this points to non-coherent variations arising from circumstellar 
material. 

On the assumption that the fundamental frequency of the group is the
rotational frequency, approximate rotational frequencies for 43 stars (i.e. 
75\,percent of the sample) can be determined.  These are listed in 
Table\,\ref{data}.  Of the remaining 14 stars, 6 are $\beta$~Cep, SPB, Maia or 
$\delta$~Sct pulsating variables.  The other stars just show low frequency 
variations which may possibly be attributed to circumstellar material.

In the 43 stars where periodicity can be found, the fundamental is usually the 
dominant peak, but the first harmonic has higher amplitude in about one-third 
of the stars.  This ratio may vary quite considerably from time to time, as 
mentioned above.

While it is always possible to attribute the quasi-periodic variations to NRP
instead of rotational modulation, this would not advance the search for the 
mass-loss mechanism since it has been shown that Be stars rotate too slowly 
for NRP to be effective in this way \citep{Zorec2016}.  A large fraction 
of non-Be stars also show rotational modulation and it is reasonable to 
suppose that Be stars are no different.  However, the typical amplitude in Be 
stars (around 1000\,ppm) is about an order of magnitude larger than in normal 
B stars (about 100\,ppm; \citealt{Balona2019c}).  This is in line with our
hypothesis (described below) in which an active region ejects gas clouds. 
In the framework presented below, it is these co-rotating gas clouds which are 
mostly responsible for the quasi-periodic light variations and not the active 
regions (starspots), though these could contribute as well.

\begin{figure}
\begin{center}
\includegraphics{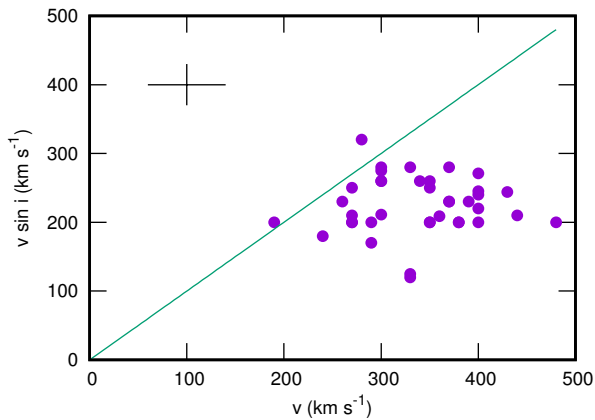}
\caption{The projected rotational velocity, $v\sin i$, as a function of the
equatorial rotational velocity, $v$, for the Be stars observed by {\it
TESS}.  The straight line represents $\sin i = 1$.  Typical error bars for
each point are shown by the cross.}
\label{vsini}
\end{center}
\end{figure}

If rotation is indeed responsible for the quasi-periodic light variations, 
then the equatorial rotational velocity, $v$, estimated from the photometric 
period and the radius, should be related to the projected rotational velocity, 
$v\sin i$.  A plot of $v\sin i$ as a function of $v$ (Fig.\,\ref{vsini}) shows 
that this is indeed the case.  Within the measured error, $v\sin i < v$ for 
all stars.  

It should be noted that the error in $v$ is considerable, as shown by the
error bars in Fig.\,\ref{vsini}.  Systematic errors in $v$ are also likely to
be present because the radius was derived from the effective temperature and
luminosity.  The derived effective temperatures for Be stars are not only 
poorly known, but probably somewhat cooler than the actual effective
temperature because it is mostly the gravity-darkened equatorial region that 
is observed.  Thus the radius is probably over-estimated, leading to larger 
$v$ for the more rapidly-rotating stars.  

Many stars are in unresolved multiple systems.  While the {\it GAIA} parallax 
will be unaffected, the apparent brightness, and hence the derived luminosity 
and radius, will be larger than that of a single star.  Thus the estimated $v$ 
will tend to be larger than its true value. On the other hand, polar 
flattening due to rapid rotation means that the estimated equatorial radius, 
and hence $v$, is likely to be too low for the most rapidly rotating stars.
The last factor probably compensates to some extent the tendency towards a
$v$ estimate which is too large.

Not all Be stars exhibit rotational modulation.  This may be due obscuration
of the photosphere.  Obscuration associated with an outburst is implied by the 
large change in apparent amplitude of the $\beta$~Cep pulsations in the 
{\it CoRoT} light curve of TIC\,234230792 (HD\,49330) and possibly also in 
27\,CMa.  

The variable projected rotational velocity of Achernar \citep{Rivinius2013b}
may be a result of gas clouds just above the photosphere.  Depending on the
latitude or height of these clouds, their velocity will vary and affect
spectral line broadening.

Because rotational modulation is common among B stars \citep{Balona2019c}, it 
is reasonable to assume that active regions are present.  This opens the way 
for a new hypothesis on the mass loss mechanism as originally proposed by 
\citet{Balona2003b} and outlined below.

\section{The mass-loss mechanism}

A strong clue to the nature of the Be phenomenon is the fact that a large
proportion of these stars have, at times, double-wave light curves (strong
first harmonic).  In the NRP interpretation, this was taken as an indication
of a preference for modes with $|m| = 2$.  The tendency for the obscuration
to be located on opposite sides of a star suggests the presence of a global
dipole magnetic field.  It is not expected that such a field is detectable
with current techniques because, as we know from the Sun and other stars,
a magnetic field of the order of one Gauss is sufficient to dominate the gas
dynamics in the outer regions of a star.

In the Sun and cool stars, flares are produced when magnetic field lines 
emanating from an active region reconnect.  Until recently, stellar flares 
were generally found only on cool K and M dwarfs, but the {\it Kepler} mission 
has led to the realization that even more energetic flares (superflares) are 
to be found in normal F and G stars \citep{Maehara2012} as well as in A and
B stars \citep{Balona2012c}.  It is thus conceivable that a flaring event 
could trigger mass loss in Be stars.  

Mass loss due to flaring is an impulsive event which describes Be outbursts 
rather well.  The highly-ionized gas released in the process will expand and be 
trapped in closed magnetic loops. If we presume that the magnetic field 
configuration is an inclined dipole, gas will be confined to a torus inclined 
to the rotational axis and concentrate at two diametrically opposite regions 
where the magnetic equator intersects the stellar equator.  This occurs because
it is in these two regions where the centrifugal force is largest.  Two 
diametrically opposed gas clouds  will account for the double-wave light curve 
in Be stars or, equivalently, the frequent occurrence of the first harmonic in 
the periodogram.  The relative strength of the fundamental and first harmonic 
depends on the relative amount of gas trapped in the two regions.

Gas will escape along open field lines and will be dragged and 
accelerated by the rapid rotation of the star. Depending on the rotational 
velocity, the gas will reach circular orbital speeds at some distance from the 
star if it is still ionized. After some time the gas, predominantly H and He, 
will cool down,  become neutral (and thus unaffected by the magnetic field) 
and slowly dissipate into the inner circumstellar disk, rotating at Keplerian 
velocity.

In this model, whether a circumstellar disk is formed or not depends on two
factors: (a) the presence of a tilted dipole magnetic field, and (b) 
sufficiently rapid rotation to reach circular velocity while the gas is still 
ionized. In the intense radiation field of an early B star, the gas will 
remain ionized to large distances.  Even if the star is rotating well below 
critical velocity, the long lever arm provided by the magnetic field will still
ensure that the gas attains circular velocity while still ionized and under 
control of the magnetic field.  This explains the wide range of rotational 
velocities in early Be stars \citep{Zorec2016}.

On the other hand, the weak radiation field in a late Be star limits the 
ionization radius closer to the photosphere. The lever arm is therefore 
shorter and requires a star already rotating close to critical for the gas to 
reach circular velocity while still ionized.  This explains why late Be stars 
are rotating in a narrow band close to the critical rotation speed.

It is evident that the condition to attain circular velocity occurs in a far 
larger range of rotational velocities in early-type stars compared to 
late-type stars.  This explains why most Be stars are of early type.  Since 
the co-rotating ionized region is much larger in early B stars, it follows 
that activity will be higher in these stars.  In this way one can understand
why early Be stars are more active than late Be stars.

\section{Conclusions}

It is clear that non-radial pulsation cannot be the trigger for mass loss
because all studies have shown that Be stars rotate well below the required
rotation rate.  In the search for an alternative mass loss mechanism,
we have examined light curves of 57 classical Be stars observed by the {\it
TESS} satellite.  

Analysis of these light curves shows that the fundamental and first harmonic 
can be identified in a large fraction (about 75\,percent) of Be stars.  
Because starspots appear to be present in normal A and B stars 
\citep{Balona2019a}, it is proposed that these regions may be the site of
magnetic re-connection events which provide the required energy for driving
mass loss with the assistance of rapid rotation.  The ejected material, 
trapped in two diametrically opposite regions by a weak tilted dipole magnetic 
field, is presumed to be the origin of the light and line profile variations
previously attributed to pulsation. 

It is found that the most common light curve (about 40\,percent) among Be 
stars is a simple single- or double-wave sinusoid.  In the periodogram this 
appears as a slightly broadened peak, or peak with fine structure, and its 
first harmonic.  The relative strengths of the first and second harmonics vary
on a timescale of months to years.  Stars showing this light curve
morphology tend to be mid- to late-B stars.

In about 30\,percent of Be stars, the periodogram still shows a fundamental 
and first harmonic, but with multiple peaks which form frequency groups.  It 
seems that none of these multiple peaks are coherent since the appearance of 
the periodogram varies on timescales of months.  These tend to be mostly mid-
to early B stars.

In about 75\,percent of Be stars, a fundamental and first harmonic can be 
identified in the periodogram.  Comparison of the derived equatorial
rotational velocity, $v$, with the projected rotational velocity, $v\sin i$,
shows that assumption of rotational modulation in these stars is consistent
with observations.

It appears that during an outburst, much of the photosphere is obscured. 
This seems to be the simplest explanation for the rapid amplitude changes
of $\beta$~Cep pulsations in HD\,49330 \citep{Huat2009} and possibly in
27\,CMa as well.

A model is proposed whereby magnetic reconnection at an active region on the 
star provides the energy for mass ejection.  In a rapidly-rotating star with
a weak global dipole field, the material may be accelerated to circular
velocity, leading to mass loss.  Other material is trapped at two
diametrically opposite regions, accounting for the light and line-profile
variations. 

Be stars are clearly very complex and no doubt exhibit a huge range of 
physical processes.  The study of the light curves, while important in
providing constraints, cannot by itself provide sufficient information to
reveal these processes.  We still do not understand the connection between the 
quasi-periodic variations which occur close to the photosphere and the 
circumstellar disc.  Further progress will require not only long-term 
photometric monitoring from space, but simultaneous spectroscopic observations 
as well.

\section*{Acknowledgments}

LAB wishes to thank the National Research Foundation of South Africa for 
financial support. Funding for the {\it TESS} mission is provided by the NASA 
Explorer Program. Funding for the {\it TESS} Asteroseismic Science Operations 
Centre is provided by the Danish National Research Foundation (Grant agreement 
no.: DNRF106), ESA PRODEX (PEA 4000119301) and Stellar Astrophysics Centre 
(SAC) at Aarhus University. 

This work has made use of data from the European Space Agency (ESA) mission 
Gaia, processed by the Gaia Data Processing and Analysis Consortium (DPAC).
Funding for the DPAC has been provided by national institutions, in particular 
the institutions participating in the Gaia Multilateral Agreement.  

This research has made use of the SIMBAD database, operated at CDS, 
Strasbourg, France.  Data were obtained from the Mikulski Archive for Space 
Telescopes (MAST).  STScI is operated by the Association of Universities for 
Research in Astronomy, Inc., under NASA contract NAS5-2655.

\bibliographystyle{mnras}
\bibliography{bestars}

\newpage

\appendix

\section*{Appendix: notes on additional stars}

\begin{figure*}
\renewcommand\thefigure{A1}
\begin{center}
\includegraphics{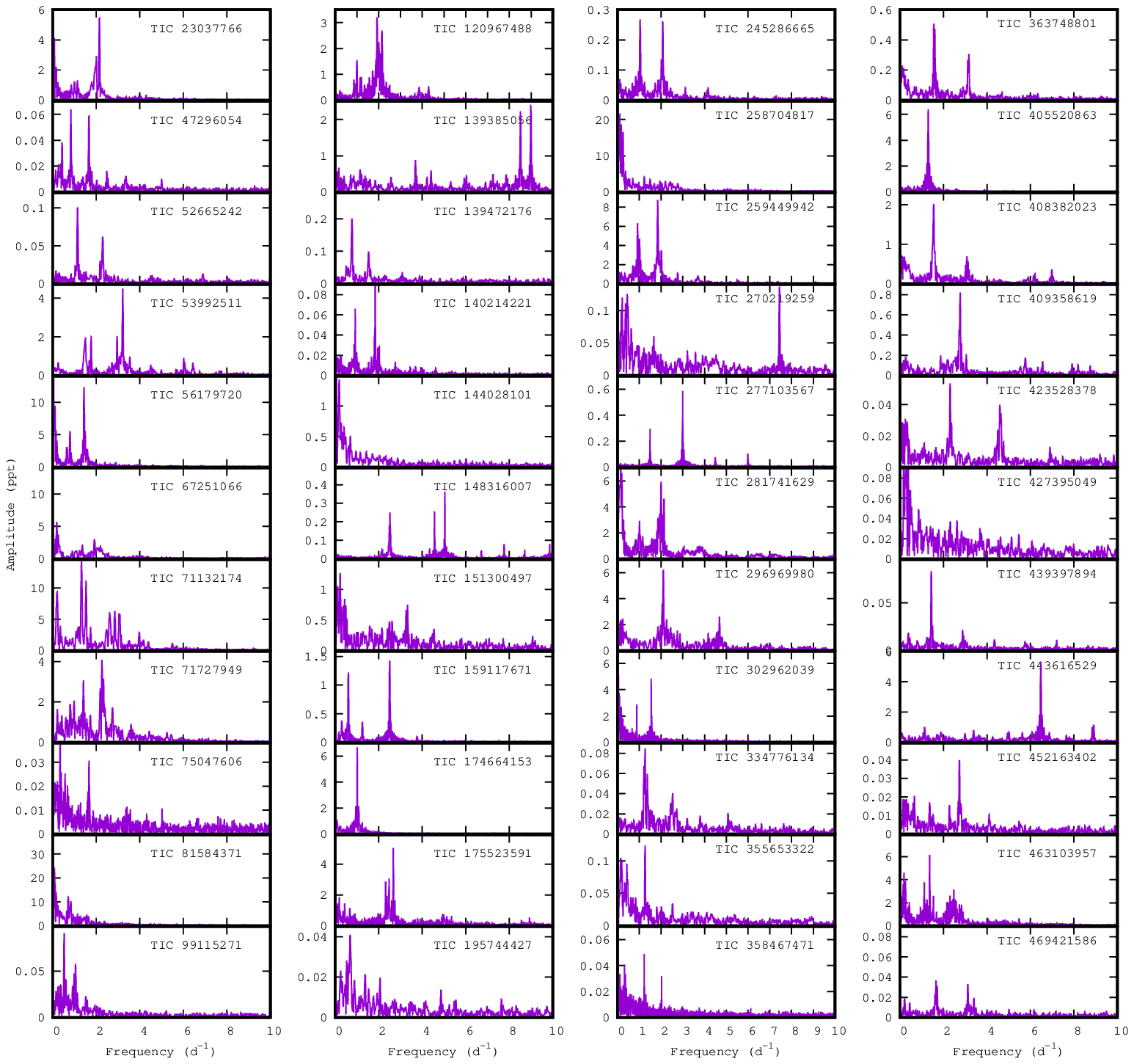}
\caption{Periodograms of stars not discussed in the main section.}
\label{short}
\end{center}
\end{figure*}

In this section notes are given on stars not discussed in the main section.
These stars were generally observed only in one sector.    Periodograms 
are shown in Fig.\,\ref{short}.  Where possible, the rotational frequency is
estimated where a fundamental and its harmonic can be discerned in the
periodogram.  This might consist of a single peak and its harmonic, a broad
peak and its harmonic or equally-spaced frequency groups each consisting of
several peaks.

\begin{description}

\item {TIC\,23037766, HR\,2787.} \citet{Balona1992b} found possible light 
variations at $1.07$ or 1.31\,d$^{-1}$ from ground-based photometry.  
The periodogram shows broad peaks at around 1.07\,d$^{-1}$ and its harmonic.

\item{TIC\,47296054, $\epsilon$~PsA.} \citet{Balona2019a} found a fundamental 
peak at 0.836\,d$^{-1}$, assumed to be the rotational frequency, and three 
harmonics as well as an anomalous peak at $0.432$\,d$^{-1}$ which is 
slightly different from half the fundamental frequency. 

\item{TIC\,52665242, HR\,2418.} The fundamental peak at $1.145$\,d$^{-1}$
and three harmonics are visible.

\item{TIC\,53992511, HR\,8408.} \citet{Cuypers1989} found a frequency of 
2.53\,d$^{-1}$ or half this value from ground-based photometry.  The 
periodogram  shows multiple peaks indicative of an SPB pulsator.  The lines 
sit on top of broad features which are repeated every 1.6\,d$^{-1}$.  This
is taken to be the rotational frequency.

\item{TIC\,56179720, 56 Eri.} From ground-based photometry, \citet{Balona1992b}
found multiple peaks at around 0.9 d$^{-1}$.  Later, \citet{Stefl1995} derived
a single photometric frequency of 0.80\,d$^{-1}$.  The periodogram shows a 
close doublet at 0.66 and 0.80\,d$^{-1}$ and a stronger peak at roughly 
double the frequency, 1.44\,$^{-1}$.  We take this to be a possible SPB
variable.

\item{TIC\,67251066, $\lambda$~Cyg.} This is a known multiple system 
\citep{Tokovinin2008}.  The periodogram shows two broad peaks at about 1.1 and 
2.2\,d$^{-1}$ which is assumed to be due to rotation.

\item{TIC\,71132174, DU~Eri, HR\,1423.} \citet{Balona1992b} found a frequency 
of 0.82\,d$^{-1}$ and its harmonic.  The periodogram shows multiple peaks in 
three frequency ranges, 1.2--1.6, 2.4--3.2 and around 4.0\,d$^{-1}$.  The 
time-frequency diagram appears to show frequency drifts, so we interpret
these three frequency groups as due to rotation at about 1.4\,d$^{-1}$.\\

\item{TIC\,71727949, HR\,2142.} This star is a Be+sdO binary system with an 
orbital period of 80.913\,d \citep{Peters2016}. \citet{Barrera1991} found a 
frequency of 1.26\,d$^{-1}$ from ground-based photometry. The main feature in 
the periodogram is a broad peak at $2.265$\,d$^{-1}$ which is variable in 
amplitude, but no rotational frequency can be determined.

\begin{table}
\renewcommand\thetable{A1}
\caption{Extracted frequencies, $\nu$ (d$^{-1}$), and amplitudes, $A$ (ppt)
for stars with high frequencies.  The figures in brackets denote the error in 
the last digit.}
\label{var}
\resizebox{\columnwidth}{!}{
\begin{tabular}{rrrrrr}
\hline
\multicolumn{1}{c}{$\nu$} & 
\multicolumn{1}{c}{$A$}   &
\multicolumn{1}{c}{$\nu$} & 
\multicolumn{1}{c}{$A$}   &
\multicolumn{1}{c}{$\nu$} &
\multicolumn{1}{c}{$A$}   \\
\hline
\multicolumn{6}{l}{TIC\,139385056.}\\
  0.1727(6) & 0.65(1) &  6.0325(8) & 0.45(1) & 12.3253(9) & 0.13(1) \\
  1.1477(7) & 0.53(1) &  7.2484(7) & 0.51(1) & 12.6811(9) & 0.09(1) \\
  2.5689(8) & 0.41(1) &  7.8538(7) & 0.49(1) & 13.0979(9) & 0.14(1) \\
  3.6910(5) & 0.86(1) &  8.5140(2) & 2.25(1) & 13.3951(7) & 0.54(1) \\
  4.2221(8) & 0.39(1) &  8.9899(3) & 2.40(1) & 14.8955(9) & 0.15(1) \\
  4.4051(7) & 0.59(1) & 10.1424(9) & 0.13(1) & 15.1680(9) & 0.09(1) \\
  5.3059(9) & 0.23(1) & 10.8241(9) & 0.31(1) & 17.9759(9) & 0.10(1) \\
  5.9381(9) & 0.35(1) & 11.8906(9) & 0.12(1) & 22.3872(9) & 0.09(1) \\
\hline
\multicolumn{6}{l}{TIC\,148316007.}\\
  2.1640(4) & 0.058(2) &  7.7650(3) & 0.082(2) & 12.3314(9) & 0.022(2) \\
  2.5229(1) & 0.238(2) &  8.7062(5) & 0.042(2) & 13.2121(9) & 0.021(2) \\
  4.3382(5) & 0.039(2) &  9.8525(3) & 0.081(2) & 16.4693(9) & 0.012(2) \\
  4.5659(1) & 0.242(2) & 10.0119(1) & 0.175(2) & 16.5653(9) & 0.013(2) \\
  5.0403(1) & 0.358(2) & 10.8473(4) & 0.050(2) & 17.8625(9) & 0.015(2) \\
  6.7145(4) & 0.045(2) & 12.1613(4) & 0.058(2) & 18.5583(9) & 0.015(2) \\
\hline                        
\multicolumn{6}{l}{TIC\,408382023.}\\
   1.5481(2) & 1.971(9) &  5.8943(9) & 0.155(5) & 10.3183(9) & 0.073(4) \\
   3.0726(5) & 0.653(7) &  6.2010(7) & 0.268(7) & 13.9655(9) & 0.114(5) \\
   3.3370(9) & 0.159(7) &  6.9841(6) & 0.358(7) &            &          \\
\hline
\multicolumn{6}{l}{TIC\,409358619.}\\
   1.8705(7) & 0.193(5) &  6.5570(9) & 0.134(5) & 11.5144(9) &  0.106(5) \\
   2.1763(9) & 0.112(5) &  7.9134(9) & 0.119(5) & 11.7959(9) &  0.104(5) \\
   2.3595(8) & 0.152(5) &  8.1915(9) & 0.101(5) & 12.3267(9) &  0.118(5) \\
   2.7678(2) & 0.812(5) &  8.7472(9) & 0.087(5) & 12.8563(9) &  0.129(5) \\
   3.0435(9) & 0.134(5) & 10.7575(9) & 0.107(5) & 14.7442(9) &  0.091(5) \\
   5.7636(7) & 0.174(5) & 10.9339(7) & 0.191(5) &            &           \\
\hline 
\multicolumn{6}{l}{TIC\,443616529.}\\
   1.1342(5) & 1.03(1) &  5.6259(8) & 0.50(1) &  8.8468(8) & 0.78(1) \\
   2.9881(8) & 0.52(1) &  6.0771(8) & 0.67(1) &  8.9063(5) & 0.91(1) \\
   3.3846(7) & 0.74(1) &  6.4739(2) & 5.25(1) & 12.1254(9) & 0.18(1) \\
   4.9365(8) & 0.51(1) &  6.8711(9) & 0.41(1) & 13.2080(9) & 0.28(1) \\
   5.0159(7) & 0.53(1) &  7.6464(9) & 0.36(1) &            &         \\
\hline 
\end{tabular}
}
\end{table}

\item{TIC\,75047606, HR\,3670.}  There is very little light variation in this 
star except for a low-amplitude peak at 1.669\,d$^{-1}$ which we take
to be the rotational frequency.

\item{TIC\,81584371, HR\,2690, FV~CMa.} This is a close optical double with
a brightness difference of 3.2\,mag.  \citet{Barrera1991} found a frequency of 
1.92\,d$^{-1}$ from ground-based photometry.  The periodogram is almost 
featureless apart from a low-amplitude broad peak at about 0.75\,d$^{-1}$ 
which is taken as the rotational frequency.

\item{TIC\,99115271, HR\,7789.} The periodogram shows a peak at 
$0.529$\,d$^{-1}$, which we assume is the rotational frequency, and its 
harmonic.

\item{TIC\,120967488, HR\,7262.} The periodogram shows three broad structures 
at frequencies 1.0, 2.0 and 4.0\,d$^{-1}$ which is assumed to be a result of
rotation at a frequency of 1.0\,d$^{-1}$.

\item{TIC\,139385056, HR\,2855, FY~CMa.} The periodogram shows high
frequencies typical of a $\beta$~Cep pulsator, but no indication of a
rotational peak.  Table\,\ref{var} lists the most significant extracted 
frequencies.

\item{TIC\,139472176, HIP\,11116.}
The periodogram shows a single peak at $0.7727$\,d$^{-1}$, assumed to be 
rotation, and several harmonics. 

\item{TIC\,140214221, HR\,1956, $\alpha$~Col.}
A peak at 0.9258\,d$^{-1}$ and its harmonic is taken as rotation.

\item{TIC\,144028101, HR\,5683, $\mu$~Lup.} A low-amplitude peak at 
33.156\,d$^{-1}$, amplitude 0.16\,ppt identifies this star as a Maia
variable.  No other significant frequency peak is present.

\item{TIC\,148316007, HR\,2507.}  There is a large number of peaks in the
range 2--20\,d$^{-1}$ indicating that this is a $\beta$~Cep star
(Table\,\ref{var}).  A broad peak at 2.523\,d$^{-1}$ and its harmonic is 
taken to be the rotational frequency.  The star lies somewhat below the
zero-age main sequence.

\item{TIC\,151300497, HR\,6397.} This is a double-lined spectroscopic
binary, with no  significant variability.

\item{TIC\,159117671, HR\,4893.}
The periodogram shows five peaks with frequencies which may be represented
by a frequency 0.3058\,d$^{-1}$ and its harmonic and 1.2522\,d$^{-1}$ and
its first two harmonics. The latter is assumed to be the rotational frequency.
The lower frequency, which corresponds to a period of 3.27\,d, may be of 
orbital nature.  The star is in a multiple system with one of the components 
being a spectroscopic binary with a period of 3.2866\,d \citep{Tokovinin2008}.

\item{TIC\,174664153, HR\,2968.}
This is a member of the open cluster NGC\,2451B.  The star appears to be a
binary in an eccentric orbit with a period of 371\,d \citep{Carrier1999}.  
There is only one significant peak in the periodogram at $1.0185$\,d$^{-1}$
which we take to be rotation.

\item{TIC\,175523591, HR\,3022.}
This star is classified as an eclipsing binary in the {\it General Catalogue 
of Variable Stars}, but no period is given. No eclipses are visible in the
{\it TESS} data.  The periodogram shows at least two groups at around
2.5\,d$^{-1}$ and its harmonic.  The first group comprises three distinct 
peaks at 2.674, 2.480 and 2.321\,d$^{-1}$, but with amplitude variations,
The mean frequency of about 2.5\,d$^{-1}$ is taken as the rotational frequency.

\item{TIC\,195744427, HR\,8028, $\nu$~Cyg.}
There are no distinctive features in the periodogram except a low-amplitude 
peak at about 0.69\,d$^{-1}$.

\item{TIC\,245286665, HR\,7719.}
The periodogram shows a presumed rotational frequency at $1.0393$\,d$^{-1}$ 
and three harmonics.

\item{TIC\,258704817, HR\,5500.}
This is a double-line spectroscopic binary \citep{Chini2012}.  The 
periodogram does not show any distinctive features.

\item{TIC\,259449942, HR\,2921.}
This star, a member of the open cluster NGC\,2422, contains a sdO companion in 
a long-period orbit \citep{Wang2018}. The periodogram shows a broad peak at 
$1.1$\,d$^{-1}$, assumed to be the rotational frequency, and several harmonics.

\item{TIC\,270219259, $\eta$~PsA.}
The star is a close double with a brightness difference of 1.09\,mag. 
\citet{Balona2019a} found broad low-frequency features at around 
$1.7$\,d$^{-1}$ for part of the time.   The most interesting feature in the
periodogram is a sharp peak at 7.445\,d$^{-1}$ (amplitude 0.139 ppt)
which must be due to pulsation.  We classify it as a Maia variable.

\item{TIC\,277103567, 29~Dor.}
From an analysis of ground-based and Hipparcos photometry, a very-low 
amplitude variation with a period of 395.48\,d was determined.  It was later
confirmed from radial velocity measurements \citep{Carrier2002}.  
\citet{Balona2019a} found a peak at 1.496\,d$^{-1}$ and three harmonics.  
This is confirmed with the additional {\it TESS} data where a peak at
$1.49668$\,d$^{-1}$, assumed to be the rotational frequency, and at least
three harmonics are visible.

\item{TIC\,281741629, JL\,212, BG~Phe.} 
This is a high Galactic latitude runaway Be star. No periodicity was found
from the first two {\it TESS} sectors \citep{Balona2019a} and no further
data are available. The possible subdwarf classification is due to
\citet{Kilkenny1989}.  The luminosity determined from the {\it Gaia} parallax 
places this star at the end of core hydrogen burning, so it cannot be a 
subdwarf.

\item{TIC\,296969980, $\theta$~Cir.}
This is a double-line spectroscopic binary \citep{Chini2012} and an 
interferometric double with an orbital period of 39.61\,yr \citep{Mason2010}.  
The periodogram consists of two peaks at 2.1047 and 4.6893\,d$^{-1}$ on 
top of broad features.  We classify it as SPB,  No rotational frequency can 
be determined.

\item{TIC\,302962039, HR\,3642.}
The periodogram features two sharp peaks at 1.5463 and 0.8875\,d$^{-1}$.  The 
light curve shows irregular variations with a timescale of about 40\,d. No 
rotational frequency can be determined.

\item{TIC\,334776134, HR\,4123.}
A broad peak at about 1.3\,d$^{-1}$ and three harmonics are visible.  We
presume this to be rotation.

\item{TIC\,355653322, $\epsilon$~Tuc.}
\citet{Balona2019a} found a low-amplitude peak at $1.2659$\,d$^{-1}$
and its harmonic.  This is assumed to be the rotation frequency.

\item{TIC\,358467471, HIP\,38779.}
This star is a member of NGC\,2516.  Two unrelated peaks are visible at 
1.2278\,d$^{-1}$ and 2.0282\,d$^{-1}$.  No rotation peak can be
identified.

\item{TIC\,363748801, $\eta$1~TrA.}
The periodogram shows peaks at 1.5548\,d$^{-1}$ (assumed to be the 
rotational frequency) and at least one harmonic.

\item{TIC\,405520863, 39 Cru, HR\,4823.}
\citet{Balona1992b} found a frequency of $1.295$\,d$^{-1}$ from ground-based
photometry.  The periodogram shows the same frequency,
$1.2982$\,d$^{-1}$, which is assumed to be due to rotation.  The 1st and 
2nd harmonics are barely visible.

\item{TIC\,408382023, HR\,3858.}
A broad peak at 1.55\,d$^{-1}$, assumed to be the rotational frequency, and two 
harmonics are visible.  In addition, some low-amplitude peaks between
5--13\,d$^{-1}$ (Table\,\ref{var}) suggest that this is a Maia variable.

\item{TIC\,409358619, HR\,5316.}
There are multiple peaks in the periodogram up to about 13\,d$^{-1}$,
indicating a $\beta$~Cep star.  Extracted frequencies are listed in 
Table\,\ref{var}.  The first harmonic of the highest-amplitude peak at 
$2.7678$\,d$^{-1}$ is faintly visible.  We assume this might be the rotational 
frequency.

\item{TIC\,423528378, $\zeta$~Crv, HR 4696.}
\citet{Barrera1991} found a photometric frequency of 1.96\,d$^{-1}$.  The 
periodogram  shows a broad peak at $2.306$\,d$^{-1}$, assumed to 
be due to rotation, as well as the 1st and 2nd harmonics.

\item{TIC\,427395049, $\theta$2~Ori, HR\,1897.}
This star is a double-line spectroscopic binary (O9.2V+B0.5V(n)) in a very
rich field \citep{MaizApellaniz2019}.  No clear periodicity is visible in 
the periodogram.

{TIC\,439397894, 2~Cet, HR\,9098.}
A peak at $1.4408$\,d$^{-1}$, assumed to be the rotational frequency, and as
many as 5 harmonics are visible.  In addition, there is a much weaker 
peak at 0.357\,d$^{-1}$ and some harmonics which may perhaps be orbital in 
nature.

\item{TIC\,443616529, $\phi$~Leo.}
Evaporation of solid, comet-like bodies grazing or falling into the star has 
been proposed as the source of variable spectral lines \citep{Eiroa2016}.
This is a $\delta$~Sct variable with several peaks (Table\,\ref{var}).

\item{TIC\,452163402, A Cen, HR\,4460.}
The strongest peak in the periodogram is at $2.7323$\,d$^{-1}$, but the 
sub-harmonic at 1.364\,d$^{-1}$ is present with a very low amplitude.
This is assumed to be the rotational frequency.

\item{TIC\,463103957, HR\,4009.}
A broad peak at 1.36\,d$^{-1}$ and its harmonic is taken as the rotation
frequency.

\item{TIC\,469421586, HR\,7843.}
This is a multiple system.  A broad peak at 1.66\,d$^{-1}$ and its harmonic
is taken as the rotational frequency.  In addition, two peaks at 
37.1229 and 41.574\,d$^{-1}$ (amplitudes 0.078 and 0.019\,ppt
respectively) suggest that it is a Maia variable.

\end{description}

\label{lastpage} 
\end{document}